\documentclass[a4paper,12pt]{article}
\pdfoutput=1
\usepackage{feynmp-auto,expdlist}
\usepackage{amsmath, amsfonts, amssymb}
\usepackage{graphicx}
\usepackage{enumerate}
\usepackage{hyperref}
\usepackage{latexsym}
\usepackage{dsfont}
\usepackage{hepnicenames}
\usepackage{enumerate}
\usepackage{soul}
\usepackage[normalem]{ulem}
\usepackage{bbold}

\newcommand{\ket}[1]{| #1 \rangle}
\newcommand{\med}[1]{\langle #1\rangle}

\usepackage{units}

\usepackage{mathrsfs,graphicx,rotating,amsmath,amsfonts,mathtools,booktabs,amssymb,wasysym}
\usepackage{hyperref}\usepackage{slashed}
\usepackage[nosort]{cite}
\usepackage[table,xcdraw,dvipsnames]{xcolor}
\usepackage{bm}
\usepackage{graphicx}
\usepackage{multirow,multicol}
\hypersetup{colorlinks,bookmarksopen,bookmarksnumbered,
linkcolor=blus,pdfstartview=FitH,urlcolor=rossos,citecolor=verde}
\allowdisplaybreaks

\renewcommand{\[}{\left[}
\renewcommand{\]}{\right]}
\renewcommand{\(}{\left(}
\renewcommand{\)}{\right)}

\renewcommand{\P}{J}
\newcommand{\Q}{\Theta}

\def\Lag{\mathscr{L}}

\newcommand{\mio}[1]{}

\def\bpm{\begin{pmatrix}}
\def\epm{\end{pmatrix}}

\usepackage{mathrsfs}

 \newcommand{\fig}[1]{~\ref{fig:#1}}
\newcommand{\sfrac}[2]{#1/#2}
 \newcommand{\One}{1\!\!\hbox{I}}

\allowdisplaybreaks
\usepackage{multicol}
\usepackage{color}
\definecolor{rosso}{cmyk}{0,1,1,0.4}
\definecolor{rossos}{cmyk}{0,1,1,0.55}
\definecolor{rossoc}{cmyk}{0,1,1,0.2}
\definecolor{blu}{cmyk}{1,1,0,0.3}
\definecolor{blus}{cmyk}{1,1,0,0.6}
\definecolor{bluc}{cmyk}{1,1,0,0.1}
\definecolor{verde}{cmyk}{0.92,0,0.59,0.25}
\definecolor{verdec}{cmyk}{0.92,0,0.59,0.15}
\definecolor{verdes}{cmyk}{0.92,0,0.59,0.4}

\newcommand{\bp}{\bar{M}_{\rm Pl}}
\newcommand{\eq}[1]{~{\rm (\ref{eq:#1})}}

\def\circa#1{\,\raise.3ex\hbox{$#1$\kern-.75em\lower1ex\hbox{$\sim$}}\,}

\newcommand{\nn}{\nonumber}
\newcommand{\beq}{\begin{equation}}
\newcommand{\eeq}{\end{equation}}

\newcommand{\bea}{\begin{eqnarray}}
\newcommand{\eea}{\end{eqnarray}}
\newcommand{\be}{\begin{equation}}
\newcommand{\ee}{\end{equation}}
\font\tenrsfs=rsfs10 at 12pt
\font\sevenrsfs=rsfs7
\font\fiversfs=rsfs5
\newfam\rsfsfam
\textfont\rsfsfam=\tenrsfs
\scriptfont\rsfsfam=\sevenrsfs
\scriptscriptfont\rsfsfam=\fiversfs

\newsavebox\MBox

\def\circa#1{\,\raise.3ex\hbox{$#1$\kern-.75em\lower1ex\hbox{$\sim$}}\,}
\makeatletter

\font\ital=cmu10

\def\hhref#1{\href{http://arxiv.org/abs/#1}{arXiv:#1}}
\usepackage{xstring}
\newcommand{\hhrefq}[1]{\IfSubStr{#1}{:}{\href{http://inspirehep.net/search?ln=en&ln=en&p=#1&of=hb&action_search=Search&sf=&so=d&rm=&rg=25&sc=0}{InSpire:#1}}{\hhref{#1}}}

\def\art{\@ifnextchar[{\eart}{\oart}}
\def\eart[#1]#2#3#4#5#6{{\rm #2}, {\em #3 \bf #4} {\rm (#6) #5} ({\em #1})}
\def\article{\@ifnextchar[{\earticle}{\oarticle}}
\def\oarticle#1#2#3#4#5#6{{\rm #1}, {\ital ``#6''}, {\rm #2 #3 (#5) #4}}
\def\earticle[#1]#2#3#4#5#6#7{{\rm #2}, {\ital ``#7''}, {\rm #3 #4 (#6) #5}  [\hhrefq{#1}]}
\def\hepart[#1]#2{{\rm #2, \sl#1}}
\def\heparticle[#1]#2#3{#2, {\ital ``#3''} [\hhrefq{#1}]}
\newcommand{\doi}[1]{\href{http://dx.doi.org/#1}{[link]}}

\newcommand{\hhrefqq}[1]{\IfBeginWith{#1}{10.}{\href{https://doi.org/#1}{doi:#1}}{\hhrefq{#1}}}
\def\earticle[#1]#2#3#4#5#6#7{{\rm #2}, {\ital ``#7''}, {\rm #3 #4 (#6) #5}  [\hhrefqq{#1}]}

\renewenvironment{thebibliography}[1]
     {\begin{multicols}{2}[\section*{\refname}]%
      \@mkboth{\MakeUppercase\refname}{\MakeUppercase\refname}%
      \list{\@biblabel{\@arabic\c@enumiv}}%
           {\settowidth\labelwidth{\@biblabel{#1}}%
            \leftmargin\labelwidth
            \advance\leftmargin\labelsep
            \@openbib@code
            \usecounter{enumiv}%
            \let\p@enumiv\@empty
            \renewcommand\theenumiv{\@arabic\c@enumiv}}%
      \sloppy
      \clubpenalty4000
      \@clubpenalty \clubpenalty
      \widowpenalty4000%
      \sfcode`\.\@m}
     {\def\@noitemerr
       {\@latex@warning{Empty `thebibliography' environment}}%
      \endlist\end{multicols}}

\newcounter{alphaequation}[equation]
\def\thealphaequation{\theequation\hbox to
0.6em{\hfil\alph{alphaequation}\hfil}}
\def\eqnsystem#1{
\def\@eqnnum{{\rm (\thealphaequation)}}
\def\@@eqncr{\let\@tempa\relax \ifcase\@eqcnt \def\@tempa{& & &} \or
  \def\@tempa{& &}\or \def\@tempa{&}\fi\@tempa
  \if@eqnsw\@eqnnum\refstepcounter{alphaequation}\fi
\global\@eqnswtrue\global\@eqcnt=0\cr}
\refstepcounter{equation} \let\@currentlabel\theequation \def\@tempb{#1}
\ifx\@tempb\empty\else\label{#1}\fi
\refstepcounter{alphaequation}
\let\@currentlabel\thealphaequation
\global\@eqnswtrue\global\@eqcnt=0 \tabskip\@centering\let\\=\@eqncr
$$\halign to \displaywidth\bgroup \@eqnsel\hskip\@centering
$\displaystyle\tabskip\z@{##}$&\global\@eqcnt\@ne
\hskip2\arraycolsep\hfil${##}$\hfil& \global\@eqcnt\tw@\hskip2\arraycolsep
$\displaystyle\tabskip\z@{##}$\hfil
\tabskip\@centering&\llap{##}\tabskip\z@\cr}
\def\endeqnsystem{\@@eqncr\egroup$$\global\@ignoretrue} \makeatother

\definecolor{Gray}{gray}{0.95}

\def\bal#1\eal{\begin{align}#1\end{align}}

\newcommand{\md}[1]{\langle #1\rangle}

\begin{document}
\vspace{1.5cm}

\begin{center}
{\LARGE\color{rossos}\bf Is negative kinetic energy meta-stable?}\\[9mm]
{\large\bf Christian Gross$^{a,b}$, Alessandro Strumia$^{a}$}\\[2mm]
{\large\bf Daniele Teresi$^{a,b}$, Matteo Zirilli$^{a}$}\\[7mm]
{\it $^a$ Dipartimento di Fisica dell'Universit{\`a} di Pisa}\\[1mm]
{\it $^b$ INFN, Sezione di Pisa, Italy}\\[1mm]
\vspace{0.5cm}

\begin{quote}\large\color{blus} 
We explore the possibility that theories with negative kinetic energy (ghosts)
can be meta-stable up to cosmologically long times.
In classical mechanics, ghosts undergo spontaneous lockdown rather than run-away
if weakly-coupled and non-resonant.
Physical examples of this phenomenon are shown.
In quantum mechanics this leads to meta-stability similar to vacuum decay.
In classical field theory, lockdown is broken by resonances and ghosts behave statistically,
drifting towards infinite entropy as no thermal equilibrium exists.
We analytically and numerically compute the run-away rate finding that it is cosmologically slow 
in 4-derivative gravity, where ghosts have gravitational interactions only.
In quantum field theory the ghost run-away rate is naively infinite in perturbation theory,
analogously to what found in early attempts to compute vacuum tunnelling; 
we do not know the true rate.
\end{quote}

\thispagestyle{empty}
\bigskip

\end{center}

\setcounter{footnote}{0}

\newpage

\tableofcontents

\section{Introduction}
A tentative quantum theory of gravity and matter is obtained writing the most generic action
with renormalizable terms, taking into account that the graviton $g_{\mu\nu}$ has mass dimension 0.
Such action is~\cite{Stelle:1976gc} 
\beq \label{eq:agravity}
S = \int d^4x \sqrt{|\det g|} \,\bigg[ \frac{R^2}{6f_0^2} + \frac{\frac13 R^2 -  R_{\mu\nu}^2}{f_2^2} -
\frac12 \bp^2 R+
\Lag_{\rm matter}
\bigg]
\eeq
where $R_{\mu\nu}$ is the Ricci tensor, $R$ is the curvature, and
$\Lag_{\rm matter}$ contains scalars, fermions and vectors.
The first two terms, suppressed by the dimension-less gravitational couplings $f_0$ and $f_2$
(in the notation of~\cite{1403.4226}),
are graviton kinetic terms with 4 derivatives.

However, a classical degree of freedom with 4 derivatives can be rewritten as 2 degrees of freedom with 2 derivatives,
and one of the two (dubbed ghost) has negative kinetic energy~\cite{Ostro}.
Gravity is no exception. The 4-derivative graviton splits 
into the massless graviton and a ghost-graviton with mass $M_2 = f_2 \bp/\sqrt{2}$.
The full action in split form can be found in~\cite{hep-th/9509142},
and the negative kinetic energy can be seen through the following simple argument.
Omitting Lorentz indices, the propagator of the 4-derivative graviton is
\beq 
\frac{1}{M^2_2 p^2 - p^4} = \frac{1}{M^2_2} \bigg[\frac{1}{p^2} - \frac{1}{p^2 - M^2_2}\bigg]
\eeq
where the minus sign indicates negative kinetic energy.\footnote{Thereby, many authors searched for a positive-energy quantization~\cite{Pais:1950za,Lee:1969fy,hep-th/0503213,astro-ph/0601672,0706.0207,1008.4678,1512.01237,1709.04925,1801.00915},
analogously to what happens for fermions (classically their kinetic
energy is undefined, but a sensible positive-energy quantum theory exists).
It is unclear what is their large-action limit 
that possibly modifies classical physics into some positive-energy version.}
It makes the theory renormalizable,
cancelling the graviton propagator at large energy $p \gg M_2$.
We explore the possibility that the degrees of freedom with 
negative kinetic-energy are physical, unlike what happens in gauge theories, where similar states
are unphysical, introduced as mathematical tools to deal with gauge redundancies.

A classical degree of freedom with positive kinetic energy interacting with negative kinetic energy 
has run-away solutions, where total energy is conserved while individual energies diverge.
Thereby negative {\em kinetic} energy is dubbed `ghost', meaning
an unphysical object to be excluded from sensible theories.
However, theories with negative and even unbounded-from-below {\em potential} energy can give sensible meta-stable physics
around a false vacuum.
Can unbounded-from-below  kinetic energy  similarly give rise to meta-stability?

To explore this issue, 
we will consider theories featuring some positive-energy degree of freedom $q_1(t)$ interacting with a ghost $q_2(t)$ 
as described by Lagrangians such as
\beq\label{eq:Ltoyq} L =
m_1\left(\frac{\dot q_1^2}{2} - \omega_1^2\frac{q_1^2}{2}\right)
\pm
m_2\left(\frac{\dot q_2^2}{2} - \omega_2^2\frac{q_2^2}{2}\right)- \frac{\lambda}{2}q_1^2 q_2^2
\eeq
as well as the analogous relativistic  theory of fields $\varphi_{1,2}(\vec x, t)$ (scalars, for simplicity)
with Lagrangian density
\be\label{eq:Lagphi12}
\Lag =\frac{(\partial_\mu \varphi_1)^2- m_1^2 \varphi_1^2}{2} 
  \pm \frac{(\partial_\mu \varphi_2)^2 - m_2^2 \varphi_2^2}{2}  -  \frac{\lambda}{2}  \varphi_1^2 \varphi_2^2  \,.
\ee
In both cases the ghost is obtained for $\pm = -1$.

We preliminarily need to address the concerns of those authors who, at this point,
dismiss the study with the motivation that an unbounded-from-below Hamiltonian is inconsistent, for example because it
allows for classical solutions that hit singularities.
These authors also view as inconsistent positive kinetic energies but with unbounded-from-below potentials.

What we want to study is how long the physical system can stay around a ``false vacuum'', before falling to other regions.  
In the case of potential meta-stability, the WKB approximation in quantum mechanics shows that
the meta-stability time is determined only by the potential barrier, 
irrespectively of the fate beyond the barrier.
The potential beyond the barrier might be unbounded-from-below (giving rise to
singular solutions) or have a true minimum: this does not affect the meta-stability time.
The fate beyond the barrier depends on possibly unknown high-energy theory.
In effective Quantum Field Theories (QFT) one considers extra non-renormalizable terms that stabilise an unbounded-from-below potential. 
As such operators have negligible impact at low field values, the meta-stability time is computable in terms of low-energy physics.

Returning back from the analogy to the argument of the present study, we want to explore
if a theory with negative kinetic energy might similarly be meta-stable up to cosmologically large times.
Let us consider, for example, the model in eq.\eq{Ltoyq}.
Its Hamiltonian is unbounded-from-below, but can be modified for example into
\beq H = \left(\frac{p_1^2}{2m_1} + m_1 \omega_1^2\frac{q_1^2}{2}\right)-\left(\frac{p_2^2}{2m_2} + m_2 \omega_2^2\frac{q_2^2}{2}\right)
+\frac{1}{2E_0}\left(\frac{p_2^2}{2m_2} + m_2 \omega_2^2\frac{q_2^2}{2}\right)^2
+ \frac{\lambda}{2}q_1^2 q_2^2\eeq
that is bounded from below and negligibly differs from the original theory at energies $E \ll E_0$ 
The energy of the 2nd degree of freedom has a Mexican-hat form that  
avoids singularities replacing them with a generalization of `ghost condensation'~\cite{hep-th/0312099}
such that $q_2$ reaches a constant but finite velocity.
The critical energy $E_0$ plays a role analogous to coefficients of non-renormalizable 
operators: in the limit where it is much higher than the energies available around the
false vacuum, it plays no role until the escape event happens.
In the following, we can thereby study the meta-stability issue in the simpler model of eq.\eq{Ltoyq} where energy is unbounded-from-below.



\smallskip

In order to see if a ghost is really excluded we start studying the problem in the simplest limit, classical mechanics.

It has been noticed that, in classical mechanics, some theories containing an interacting ghost have
stable classical solutions with appropriate initial conditions dubbed  ``islands of stability''~\cite{Narnhofer:1978sw,Pagani:1987ue,hep-th/0407231,1302.5257,1607.06589,1703.08929,1811.07733,1811.10019,1902.09557,2003.10860}. This happens even when interactions are generic enough that
no constant of motion forbids interacting ghosts to evolve towards catastrophic run-away instabilities.
Rather, ghosts undergo spontaneous lockdown, with energies that vary but remain in a non-trivial restricted range.
Studies based on numerical computations of classical time evolution 
cannot reach cosmological meta-stability times, so an 
analytic understanding is needed.
Extending earlier works~\cite{Pagani:1987ue} we will show that the needed mathematics had been already developed
to understand a related problem: why the solar  system is meta-stable, despite that no constant of motion forbids planets to escape?
Oversimplifying, it has been shown that
classical systems that can be approximated as oscillators plus {\em small} interactions
tend to undergo ordered epicycle-like motions, while large interactions lead to chaos.
We will see that this implies that ghosts with large interactions run away, but
ghosts with generic {\em small} interactions are stable.
Weakly coupled theories contain hidden quasi-constants of motion.
Since this might appear exotic, in Appendix~\ref{mainstream} we recall that known physical systems exhibit this behaviour:
asteroids around the Lagrangian point $L_4$ and electrons in magnetic fields plus repulsive potentials
are described by a ghost degree of freedom, and yet they are meta-stable.

Since classical mechanics does not exclude ghosts, 
in section~\ref{QM} we study quantum mechanics, finding that meta-stability persists:
a ghost (negative kinetic energy, $K$-instability) is not qualitatively less meta-stable 
than a negative potential energy ($V$-instability). 

However, resonances (such as $\omega_1=\omega_2$ in eq.\eq{Ltoyq}) can lead to ghost run-away even at small
coupling, depending on the specific form of the interaction.
Studying in section~\ref{CFT} 
classical field theory we encounter an infinite number of resonances, by expanding a field in Fourier modes.
While local field theories can give resonances of benign type, the infinite number of resonances 
removes the hidden constants of motion.
We then perform a statistical analysis showing that systems containing ghosts
do not have a thermal state: heat keeps flowing from ghost fields to positive-energy fields, because this increases
entropy. We compute the rate of this instability through Boltzmann equations, 
finding a rate not exponentially suppressed by small couplings.
Nevertheless, in the special case of 4-derivative gravity,
the graviton ghost has Planck-suppressed interactions which are small enough that the
ghost run-away rate is not problematic in cosmology.
We validate this analytic understanding through classical lattice simulations.

\medskip

In section~\ref{QFT} we finally consider relativistic quantum field theory, which is the relevant but most difficult theory.
By performing the zero-temperature limit of Boltzmann equations we find a divergent tree level ghost run-away rate.
Such divergence arises because the  initial vacuum state is Lorentz-invariant, giving rise to an integral over
the non-compact Lorentz group that describes a boost of the final state.
The same Lorentz integral arose in earlier computations of $V$-instability tunnelling,
but Coleman later argued that that vacuum decay can be computed in terms of a 
Lorentz-invariant instanton, the `bounce', and its rate is exponentially suppressed at small coupling.
We don't know if something similar holds for $K$-instability.

Conclusions are presented in section~\ref{concl}.

\section{Ghost meta-stability in classical mechanics?}\label{class}
We consider a degree of freedom $q(t)$ in 0+1 dimensions with 4-derivative kinetic term
\beq L =-\frac{1}{2} q \bigg(\frac{\partial^2}{\partial t^2} +\omega_1^2\bigg) \bigg(\frac{\partial^2}{\partial t^2}+\omega_2^2\bigg)q - V_I (q,\ddot q)\eeq
where the first term is quadratic in $q$ and $V_I$ contains interactions.
We add zero as a perfect square containing an auxiliary degree of freedom $\tilde q$ with no kinetic term:
\beq \label{eq:adda}
L =\frac 12 \bigg[- \ddot q^2 +(\omega_1^2+\omega_2^2) \dot q^2 - \omega_1^2\omega_2^2 q^2\bigg]+
\frac12 \bigg[ \ddot q +  (\omega_1^2+\omega_2^2)\frac{q}{2} - \frac{\tilde q}{2}\bigg]^2 -V_I.\eeq
Expanding the square cancels both the second-order and the fourth-order kinetic terms leaving
\beq L =-\frac{\tilde q \ddot q}{2} + (\omega_1^2-\omega_2^2)^2\frac{q^2}{8}  - (\omega_1^2+\omega_2^2)\frac{\tilde qq}{4} + \frac{\tilde q^2}{8} -V_I.\eeq
The kinetic and mass terms are diagonalised performing the field redefinition
\beq \left\{\begin{array}{l}
\tilde q = \sqrt{\omega_2^2 - \omega_1^2}(q_1 - q_2)\cr
q = (q_1 + q_2)/{\sqrt{\omega_2^2-\omega_1^2}}
\end{array}\right.
\eeq
obtaining, after an integration by parts
\beq L = \frac{\dot q_1^2 - \omega_1^2 q_1^2}{2} -  \frac{\dot q_2^2 - \omega_2^2 q_2^2}{2}  -V_I\left(
\frac{q_1+q_2}{\sqrt{\omega_2^2-\omega_1^2}},-\frac{\omega_1^2 q_1 + \omega_2^2 q_2}{\sqrt{\omega_2^2 - \omega_1^2}}
\right).  \eeq
We can thereby focus on the toy model of eq.\eq{Ltoyq}
that captures the relevant physics.
This classical theory only has one free physical parameter, $\omega_1/\omega_2$, plus the initial conditions for its time evolution.
Indeed, without loss of generality we can rescale $q_1$ and $q_2$ to set $m_1 = m_2 =1$.
By rescaling $t$ we can set $\omega_1=1$.
Furthermore, classical physics is invariant under a multiplicative rescaling of $L$,
so that we could set $\lambda=1$. 
To improve readability we keep $\omega_1$, $\omega_2$ and $\lambda$ as apparent parameters, 
but it should be clear that our following analysis is general.


The classical equations of motion are
\beq \ddot q_1 + \omega_1^2 q_1 +\lambda q_1 q_2^2 = 0,\qquad
\ddot q_2 + \omega_2^2 q_2 -\lambda q_2 q_1^2 = 0.\eeq
The only constant of motion is the total energy $E = E_1 - E_2 + V_I$, which is conserved, where
\beq E_i = \frac{\dot q_i^2}{2} +\omega_i^2\frac{q_i^2}{2}>0,\qquad
V_I = \frac{\lambda}{2} q_1^2 q_2^2\eeq
while $E_1$ and $E_2$ are not conserved, e.g.\ $\dot E_1 = - \lambda q_2^2 d(q_1^2)/dt$.
No conservation law prevents rapid ghost run-away to $E_1,E_2\to \infty$.
Numerical evolution shows that solutions starting from $|E_1 - E_2| \circa{>} V_I$ quickly undergo run-away.
On the other hand, for solutions starting from small enough initial energies $E_1, E_2 \ll V_I$,
$E_1(t)$ and $E_2(t)$ evolve remaining 
confined to a small range, for a time longer
than what can be numerically computed.\footnote{In agravity, this kind of initial conditions correspond to
small gradients, that might be selected by inflationary cosmology~\cite{1902.09557}.}
Analytic work is needed to understand this surprising phenomenon.

\subsection{Action-angle variables}
A technique used to study perturbed quasi-periodic motions in celestial mechanics is useful.
Considering one pair $(q,p)$ of Hamiltonian variables,
it is useful to pass to canonical action-angle variables $(\Q,\P )$ such that the Hamiltonian only depends on $\P$
and motion is immediately solved.

In the simplest case of an harmonic oscillator, this gives
\beq H =\frac{p^2}{2m} +\frac{m\omega^2 }{2} q^2 = \omega \P \eeq
where $m>0$ ($m<0$) for a normal particle (a ghost).
The canonical transformation is
\beq \label{eq:harmqp}
q  =\sqrt{\frac{2\P }{m\omega}}\sin \Q,\qquad
p = \sqrt{2m \omega \P }\cos \Q\eeq
and its inverse is
\beq \Q =\arccos\frac{p}{\sqrt{p^2+(m \omega q)^2}} ,\qquad
\P  = \frac{p^2+(m \omega q)^2}{2m\omega}.\eeq
One can verify that $[\Q,\P ] = (\partial \Q/\partial q)(\partial \P /\partial p)-(\partial Q/\partial p)( \partial\P /\partial q)=1$ 
or more formally write the generator of the canonical transformation
\beq \label{eq:WqJharm}
W(q,J)=\int p\, dq =
\frac12 q \sqrt{m\omega(2\P-m\omega q^2)} + \P \arccos\sqrt{1-\frac{m\omega q^2}{2\P}} .
\eeq
In action-angle variables $H = \omega \P $ so that motion of a harmonic oscillator
is trivially solved by $ \Q = \Q_0+\omega t$, $\P  = E/\omega$.
For a generic anharmonic oscillator, the transformation 
to action-angle variables such that $H$ depends only on $\P_i $
cannot
be written analytically.

\begin{figure}[t]
$$\includegraphics[width=0.4\textwidth]{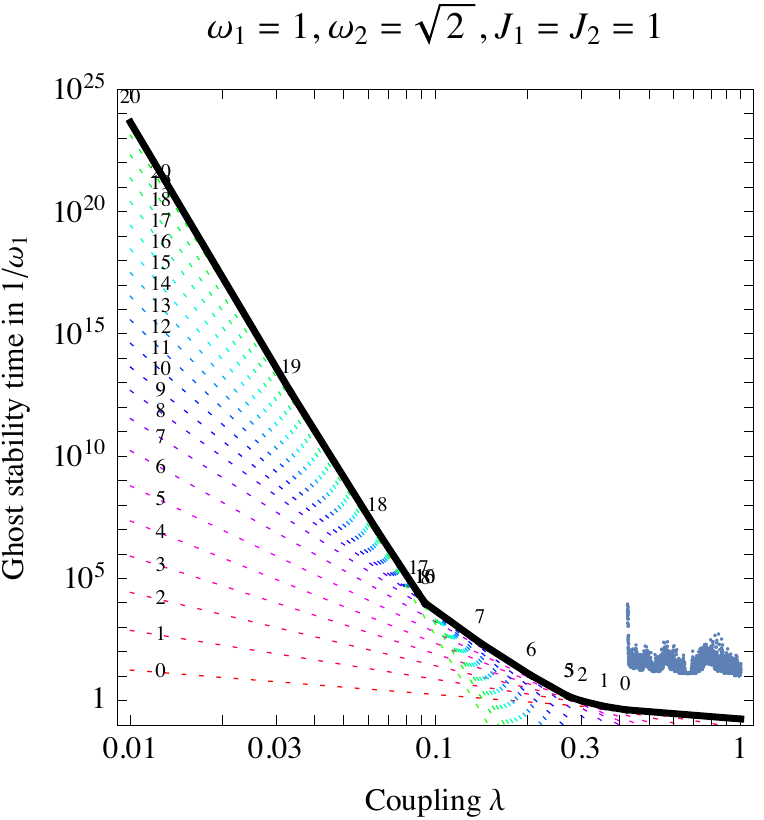}\qquad
\includegraphics[width=0.4\textwidth]{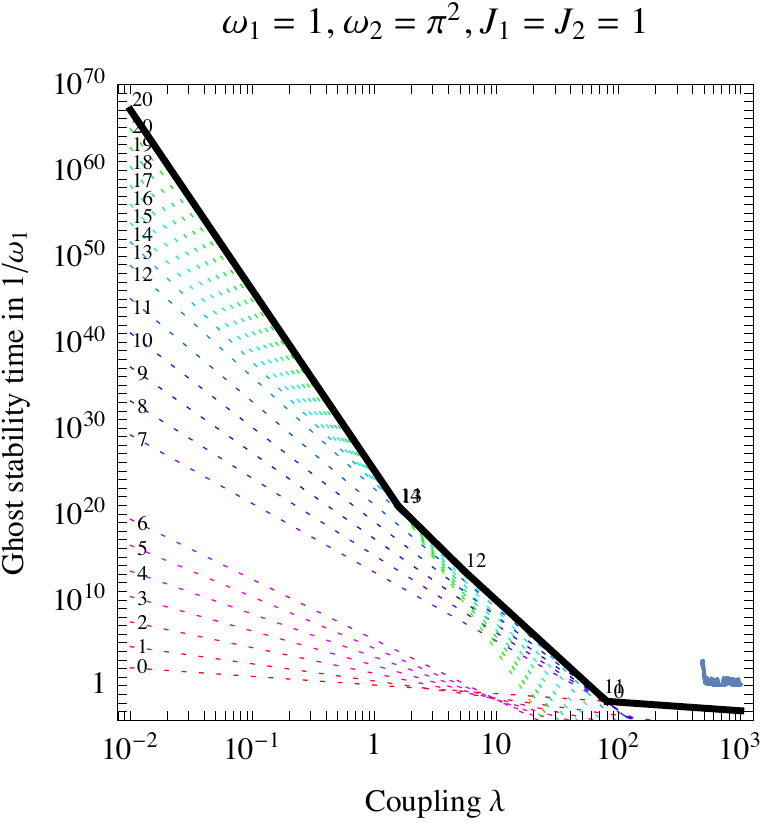}$$
\caption{\em We consider the ghost model of eq.\eq{Ltoyq} with $n=2$ degrees of freedom
and quartic coupling $\lambda$.
The dots are numerical results of observed ghost instability.
The black curve is the analytic lower bound on the ghost stability time, computed up to 20th order in $\lambda$.
\label{fig:StabilityBound}}
\end{figure}

Going to action-angle variables for the two free harmonic oscillators, our toy ghost model of eq.\eq{Ltoyq} becomes
\beq \label{eq:Hq1q2}
H = \omega_1 \P _1 - \omega_2 \P _2 + \epsilon \P _1 \P _2 \sin^2 \Q_1 \sin^2 \Q_2\qquad\hbox{where}\qquad
\epsilon =   \frac{2 \lambda}{\omega_1 \omega_2}
\eeq
and $E_i = \omega_i J_i \ge0$.
The $-$ signals a ghost.
The change of variables makes numerics  stable up to longer time scales.
Starting from $t=0$, fig.\fig{StabilityBound} shows the time $t_{\rm end}$ at which the ghost
run-away happens as function of $\lambda$ for some fixed initial conditions and
given $\omega_2/\omega_1$.
We see a chaotic behaviour at larger $\lambda$ that sharply starts above some critical value.

\begin{figure}[t]
$$\includegraphics[width=0.42\textwidth]{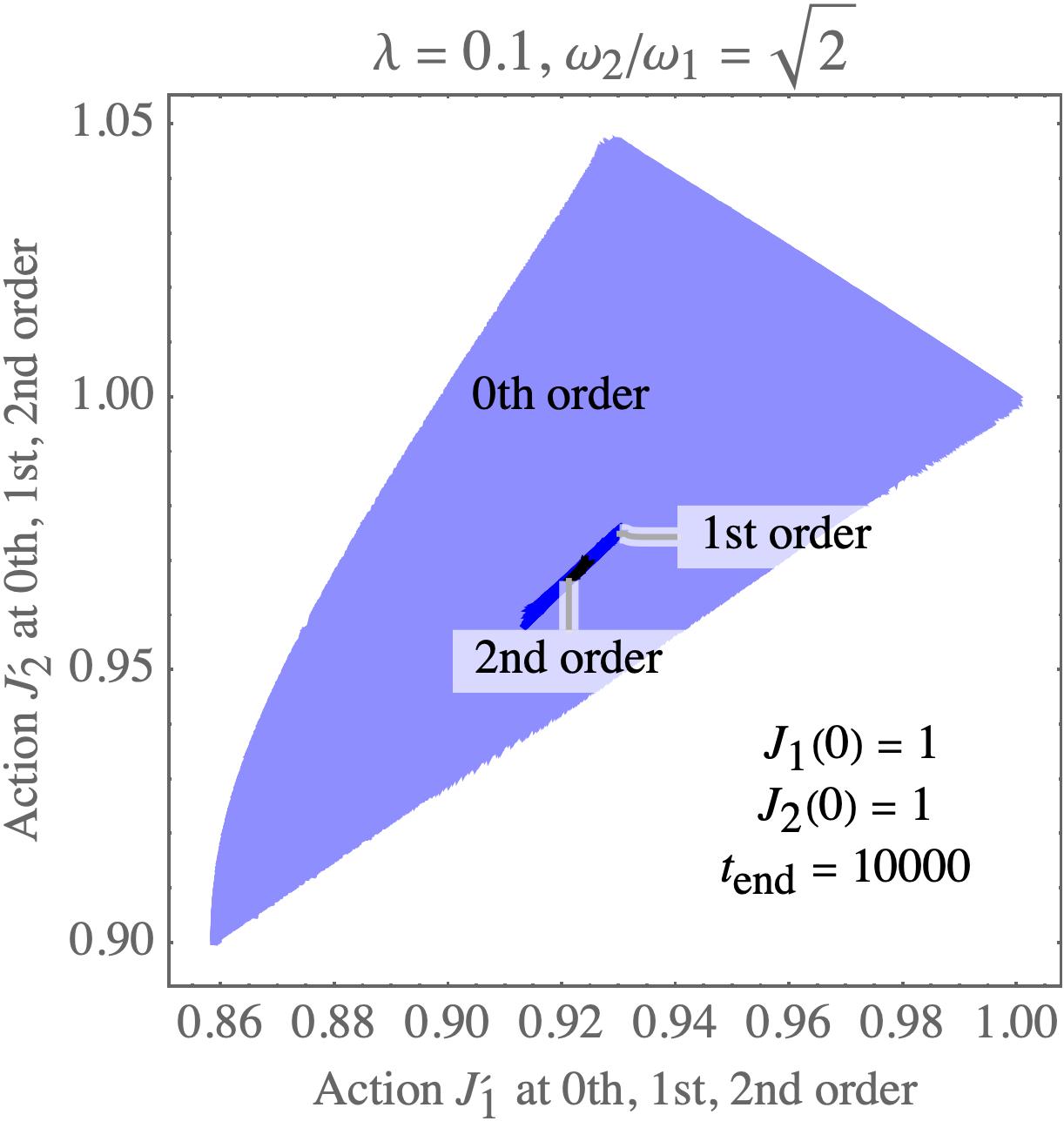}\quad
\includegraphics[width=0.42\textwidth]{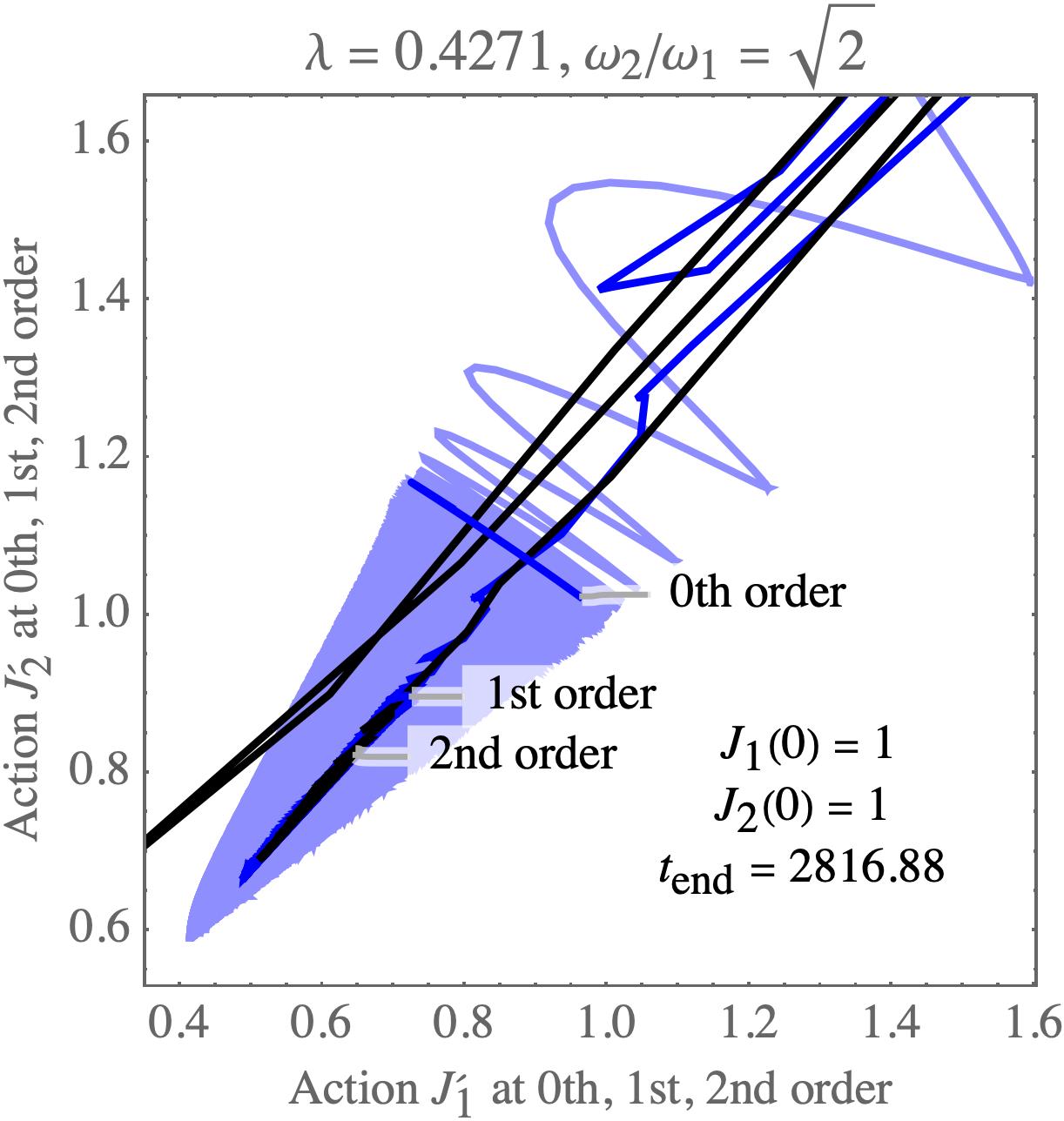}$$
\caption{\em Time evolution of the two quasi-conserved energies $(\P '_1,\P '_2)$
computed at 0th-order (lighter, $\P '_i=\P _i$),
1st-order (medium), 2nd order (darker) in $\lambda$.
For small $\lambda$ (left plot) time evolutions remains in a confined region that gets smaller and smaller
as higher orders are included.
Meta-stability is lost above some critical value of the coupling $\lambda$ (right plot), when
the Birkhoff series in $\lambda$ stops converging.
%
\label{fig:BetterCte}}
\end{figure}

Fig.\fig{BetterCte}a shows that, for small $\lambda$,
$\P _1$ and $\P _2$ remain confined in a well-defined region up to long times,
while $\Q_1$ and $\Q_2$ evolve almost linearly in time.
Analytic work is needed to know if smaller $\lambda$ leads to meta-stability or to absolute stability.
The region in the $(\P _1,\P _2)$ plane extends with increasing $\lambda$ until suddenly chaos and ghost run-away take over.


\medskip

This behaviour is characteristic of near-integrable system.
Integrable systems (such as $n$ independent oscillators)
are those for which any trajectory evolves along tori in phase space,
rather than filling higher-dimensional sub-spaces up to the whole phase space.
Adding small interactions, 
a near-ordered behaviour persists because the system can be computed perturbatively.
In the case of ghosts, this implies their meta-stability.
For large coupling the perturbative expansion fails and the system becomes chaotic.
If the system contains ghosts, this leads to run-aways.

For small $\epsilon$ we can analytically solve the equations of motion as power series in $\epsilon$.
At 0th order in $\epsilon=2\lambda/\omega_1\omega_2$ the equations of motion are solved by
\beq \P _i(t) =\P _{i0},\qquad \Q_1(t) =\Q_1(0) + \omega_1 t,\qquad \Q_2(t)=\Q_2(0)-\omega_2 t.\eeq
We see that $\P _i(t) = \P _{i0}$ are constant, for both $i=\{1,2\}$.
Their equations of motion at 1st order
\beq \P '_1 = -\epsilon \P _{10} \P _{20} \sin (2\omega_1 t)\sin(\omega_2 t)^2,\qquad
\P '_2 = \epsilon \P _{10} \P _{20} \sin (2\omega_2 t)\sin(\omega_1 t)^2\eeq
are solved by
\beq \label{eq:P11storder}
\P _1(t) = \P _{10} + \epsilon \P _{10}\P _{20}
 \frac{\omega _1 \omega _{2-1} \cos (2 t \omega _{1+2})+\omega _{1+2} \left(2
\omega _{1-2} \cos \left(2 t \omega _1\right)-\omega _1 \cos (2 t \omega _{1-2}\right))+2 \omega _2^2}{8\omega_1 (\omega_1 - \omega_2)(\omega_1 + \omega_2)}\eeq
having defined $\omega_{1-2}=\omega_1-\omega_2$ etc.
The dimension-less expansion parameter is $\sim \epsilon J/\omega$, that describes
the energy in the interaction term divided by the energy in the
free quadratic part of the Hamiltonian.
This 1st  order approximation fails after some oscillations; 
nevertheless for small $\epsilon$ it approximates well the range of $(\P _1,\P _2)$ covered by the full numerical solution.
The 1st order perturbation diverges if $\omega_1=\pm \omega_2$.
More in general, higher orders diverge if $\omega_1$ and $\omega_2$ are
`commensurable', namely if  the resonance condition
$N_1\omega_1 + N_2\omega_2=0$ is satisfied for some  integers $N_{1,2}$.\footnote{These resonances correspond to what in field theory are on-shell scattering and decay processes:
in zero spatial dimensions $\omega_i$ do not depend on momenta, 
so that on-shell processes are only possible among appropriate integer numbers $N_i$ of modes.}

A more important problem is that the perturbative series in $\epsilon$ (or $\lambda$) is not convergent 
but at most asymptotic.
Thereby its existence does not imply absolute stability and a more complicated analysis is needed, yielding stability over exponentially long times.

\subsection{Perturbative Birkhoff series}
We consider the toy model described by the Hamiltonian of eq.\eq{Hq1q2}.
Rather than finding solutions perturbatively in $\epsilon$ we follow a more general, equivalent, approach.
We seek to `diagonalise' the classical Hamiltonian.
Namely, we search for a canonical transformation $\P _i \to \P '_i$ and $\Q_i \to \Q'_i$ such that 
the Hamiltonian does not depend on $\Q'_i$:
\beq H(\P _i, \Q_i) = H'(\P '_i).\eeq
We perform a generic canonical transformation with generator
\beq  \label{eq:can}
\P '_i \Q_i +W(\P ',\Q)  \qquad\hbox{i.e.}\qquad
\P  = \P ' + \partial_{\Q_i} W,\qquad \Q'=\Q+\partial_{\P '} W.\eeq
So, defining 
$ f= \sin^2 \Q_1 \sin^2 \Q_2$ one gets
\beq \label{eq:Hbir}
H'(\P ') = H(\P ) = \omega_1 (\P '_1 + \partial_{\Q_1} W)-\omega_2 (\P '_2 + \partial_{\Q_2} W)
 + \epsilon f (\P '_1 + \partial_{\Q_1} W)(\P '_2 + \partial_{\Q_2} W).\eeq
If we could solve this equation, all $\P '_i$ would be exact constants of motion and the system would be integrable.
However, we can only expand and perturbatively solve eq.\eq{Hbir} in powers of $\epsilon$,
 \beq W = \epsilon W^{(1)} +\epsilon^2  W^{(2)}+\cdots,\qquad
 H' = H + \epsilon H^{(1)}+\epsilon^2 H^{(2)}+ \cdots.\eeq
Since the system is not integrable, the Birkhoff series is only asymptotic
and $\P '_i$ are the approximated constants of motion
observed in numerics.
Because of the periodicity in $\vec \Q \equiv (\Q_1,\Q_2)$, we expand each term in Fourier series, e.g.
\be 
W^{(n)}(\Q_1,\Q_2) = -i \sum_{N_1,N_2 = - \infty}^\infty e^{i \vec N \cdot \vec \Q} \, W^{(n)}_{\vec N} ,\qquad
\vec N = (N_1,N_2).
\ee
The only non-zero coefficient of the Fourier series of
$f = \sum_{\vec N} e^{i N_i \Q_i} f_{\vec N}$
are $f_{00} = 1/4$, $f_{\pm 2, \pm 2}=f_{\pm 2, \mp 2} =1/16$, $f_{\pm 2,0}=f_{0,\pm 2} = -1/8$.

\subsubsection{First order in the coupling}
Expanding eq.\eq{Hbir} at first order gives
\beq  \label{eq:H1model}
H^{(1)}=  \omega_1\frac{\partial W^{(1)}}{\partial \Q_1}-\omega_2 \frac{\partial W^{(1)}}{\partial \Q_2} + \P '_1 \P '_2 f(\Q_1,\Q_2) .\eeq
The first term involves derivatives of a periodic function with period $2 \pi$, by the very definition of the
angle variables $\Q_i$. Therefore, its average over a period is zero. Averaging over $\Q_i$ we get
\be 
H^{(1)} = \frac{\P _1' \P _2'}{4}\qquad \hbox{i.e.}\qquad H' = \omega_1 \P '_1 - \omega_2 \P '_2 + \epsilon \frac{\P _1' \P _2'}{4}+{\cal O}(\epsilon^2).
\ee
We next compute the canonical transformation $W^{(1)}$ through the Fourier expansion.
We  get $W^{(1)}_{00}=0$ and 
\be \label{eq:W1model}
W^{(1)}_{N_1 N_2} = \frac{\P _1' \P _2' f_{N_1 N_2}}{ N_2 \omega_2-N_1 \omega_1 } \,
\ee
for $N_1,N_2 \neq 0$. Summing over the non-vanishing $\vec N$ this means
\beq 
W^{(1)}=\frac{\P '_1 \P '_2}{8(\omega_2^2-\omega_1^2)}\left[
(\omega_1 \cos(2\Q_2)-\omega_1 + \frac{ \omega_2^2}{\omega_1})\sin(2\Q_1)+
(\omega_2\cos(2\Q_1)-\omega_2+\frac{\omega_1^2}{\omega_2})\sin(2\Q_2)\right]
\eeq
and thereby
\be 
\P _1' = \P _1 + \epsilon \frac{\P _1 \P _2}{4 \omega_1 (\omega_1^2 - \omega_2^2)} \[ \cos 2 \Q_1 \( \omega_2^2 - \omega_1^2 + \omega_1^2 \cos 2 \Q_2 \) - \omega_1 \omega_2 \sin 2 \Q_1 \sin 2 \Q_2\] + {\cal O}(\epsilon^2)
\ee
which gives the extra approximate integral of motion (in addition to energy, an exact constant).
At this order the only resonance is $\omega_1 = \pm \omega_2$.
The perturbative expansion fails close to the resonance.
The numerical solution shows that $\P '_1$ is an approximate pseudo-integral of motion
for small $\epsilon$, unless $ \omega_1 \approx \omega_2$.

 
\subsubsection{Generic order in the coupling} 
Eq.~\eqref{eq:Hbir} expanded at order $n>1$ ($n=1$ is special) is
\begin{eqnarray}
H^{(n)}(\P '_1,\P '_2) &=& \omega_1 \frac{\partial W^{(n)} }{\partial \Q_1} - \omega_2 \frac{\partial W^{(n)} }{\partial \Q_2} + \; f(\Q_1,\Q_2) \times \notag\\
&& \times\[\P _1' \frac{\partial W^{(n-1)}}{\partial \Q_2}  + \P _2' \frac{\partial W^{(n-1)}}{\partial \Q_1}  + 
\sum_{m=1}^{n-2} \frac{\partial W^{(m)} }{\partial \Q_1}  \frac{\partial W^{(n-1-m)}}{\partial \Q_2} \].
\end{eqnarray}
At each order $n$ only a finite set of coefficients of $W^{(n)}_{N_1 N_2}$
are non-zero, since $f$ only has few non-zero Fourier coefficients. 
The constant term ($ p_{1}=p_2= 0$) 
allows to find explicitly the Hamiltonian, whereas the other terms give the canonical transformation. We may fix the freedom of performing $\Q$-only transformations by choosing $W^{(n)}_{0 0} = 0$ finding 
\begin{align}
H^{(n)} &= \sum_{\vec q + \vec r = \vec 0}  (r_2 \P _1' + r_1 \P _2') f_{\vec q} W_{\vec r}^{(n-1)}  +  
\sum_{m=1}^{n-2}   \sum_{\vec q + \vec r + \vec s = \vec 0} r_1 s_2 f_{\vec q} W_{\vec r}^{(m)} W_{\vec s}^{(n-1-m)} , \\
W^{(n)}_{\vec N} &= \frac{1}{N_2 \omega_2-N_1 \omega_1}\[\sum_{\vec q + \vec r = \vec p} (r_2 \P _1' + r_1 \P _2') f_{\vec q} W_{\vec r}^{(n-1)}  +
\sum_{m=1}^{n-2}  \sum_{\vec q + \vec r + \vec s = \vec p} r_1 s_2 f_{\vec q} 
W_{\vec r}^{(m)} W_{\vec s}^{(n-1-m)} \],
\end{align}
which are explicit equations for the Hamiltonian and the canonical transformation at order $n$ in terms of the lower orders.
 $H^{(n)}$ is a polynomial of degree $n+1$ in $\P '_{1,2}$ with coefficients that depend on $\omega_i$.\footnote{Relations such as
 $W^{(n)}_{-N_1, N_2}(\omega_1,\omega_2) =(-1)^n W^{(n)}_{N_1, N_2}(-\omega_1,\omega_2) $
 allow to compute only for positive $N_{1,2}\ge 0$, if $\omega_i$ are left generic.
 However this produces cumbersome expressions, and 
computations are more efficiently performed setting $\omega_i$ to numerical values, such that each
term is a short polynomial in $\P '_i$.}

\subsection{Stability estimates}\label{sec:stability_estimates}
For typical interacting systems, frequencies vary depending on initial conditions and can thereby hit resonances,
invalidating the Birkhoff series that guarantees stability.
Kolmogorov proved that instability only happens for a sub-set of values of initial conditions
that are as rare as rational numbers within real numbers:
most initial conditions lead to stable motion.
For systems with 2 degrees of freedom and conserved energy this is enough to guarantee exact stability,
because there is only one quasi-constant of motion, say $\P '_1$.
Any initial condition is `surrounded' by nearby values so that stability holds.
On the other hand, with more than 2 degrees of freedom there are 2 or more quasi-constants $\P '_i$,
so that they can undergo Arnold diffusion:
their values slowly drift through the rare instabilities, not being surrounded by stable values.
This drift is not visible in perturbation theory because it takes place for `rational' values of $\omega_i$
such that perturbation theory fails. 
Nekhoroshev estimated that the drift is non-perturbatively slow, giving rise to an exponentially large
instability time~\cite{Nek}.

In concrete systems the meta-stability time can be computed as follows.
The perturbative Birkhoff series allows to remove interactions up to an arbitrarily large power $\epsilon^k$
so that the remaining small interaction 
can destroy stability on long enough time-scales, of order $\epsilon^{-k}$.
As the Birkhoff series is only asymptotic, stability estimates are obtained by computing up to  some high optimal order
in the asymptotic expansion.
For example,~\cite{Giorgilli} computed the meta-stability time of asteroids around the Lagrangian point L4,
that contain a ghost degree of freedom.



\smallskip

In our model, we can compute
the time $\tau_n(\P _{\rm max} ^{\rm in}\to \P _{\rm max} )$ for which we are guaranteed that any evolution 
starting from $\P '_i \leq  \P ^{\rm in}_{\rm max}$
remains within $\P '_i \leq  \P _{\rm max} > \P _{\rm max}^{\rm in} $.
We maximise over $\P _{\rm max}$, when possible, having in mind Lyapunov stability, so that $\tau_n(\P _{\rm max}^{\rm in}) \equiv \max_{\P _{\rm max}} \tau_n(\P _{\rm max} ^{\rm in}\to \P _{\rm max}  )$. 

Computing at different orders $n$ in the expansion give different $\P '_i$ and different times $\tau_n$; 
because of the asymptotic character of the Birkhoff series stability is guaranteed up to the largest $\tau_n$.
Non-conservation of $\P '_i$ happens because interactions $\delta H$ remain at higher order:
\beq H(\P _i,\Q_i) = H^{(\le n)}(\P '_i) +\delta H (\P '_i, \Q_i) \eeq
where $H^{(\le n)} = \sum_{k=0}^n \epsilon^k H^{(k)}$ includes terms up to order $n$.
The leading-order contribution to the residual is
\beq \delta H = -  \epsilon^{n+1} \left[
 \omega_1 \frac{\partial W^{(n+1)} }{\partial \Q_1} - \omega_2 \frac{\partial W^{(n+1)} }{\partial \Q_2}\right]+{\cal O}(\epsilon^{n+2}) .
 \eeq
Such term can be computed from its Fourier coefficients
\be 
\delta H^{(n+1)}_{\vec N}
= 
\left\{ 
\begin{array}{ll}
H^{(n+1)} \qquad &\hbox{for }\vec{n} = \vec{0} \\
(N_2 \omega_2 - N_1 \omega_1) W_{\vec N}^{(n+1)}  \qquad &\hbox{for } \vec{n} \neq \vec{0}
\end{array} \right.  .
\ee
The residual time evolution of $\P '_i$ is given by its Hamiltonian equation of motion
\beq   \dot \P '_i = - \frac{\partial}{\partial \Q'_i} \delta H
\eeq
where, at leading order in the residual, we can approximate 
$ {\partial}/{\partial \Q'_i} \simeq \sfrac{\partial}{\partial \Q_i}$ and thereby
avoid re-expressing $\Q$ in terms of $\Q'$ in $\delta H$.
A lower bound on the stability time is obtained by substituting $\dot \P '_i $ with its maximal value.
Neglecting higher orders in $\epsilon$:
\be \label{eq:boundlead}
\left| \frac{\partial}{\partial \Q'_i} \delta H \right| \leq 
\sum_{\vec p} \left| \frac{\partial}{\partial \Q'_i} \delta H^{(n)}_{\vec p} \right|=
\sum_{N_1,N_2} \left| N_i (N_2 \omega_2 - N_1 \omega_1)  W_{\vec N}^{(n)} \right|
\ee
having used the triangular inequality. 
Higher orders in $\epsilon$ weaken the bound in eq.\eq{boundlead} by a factor of 2~\cite{Giorgilli}. 


\subsubsection{Stability at lowest order}
To start, we outline the procedure at lowest order, such that
the approximately conserved quantities are simply $\P '_i = \P _i$ 
and the remainder in the Birkhoff series simply is the whole interaction
\be 
\delta H^{(0)} = 2 \lambda  \frac{\P _1 \P _2}{\omega_1 \omega_2} \sin^2 \Q_1 \sin^2 \Q_2.
\ee 
To compute the stability time, we use the inequality
\be 
|\P '_i(t) - \P '_i(0)| \leq t \max_{\P '_i \leq \P _{\rm max}} \left| \dot \P '_i \right |  \le t \,  \frac{2\lambda \P _{\rm max}^2}{\omega_1\omega_2} .
\ee
The region can be abandoned only after a time
\be 
t \geq  \tau_{0}(\P ^{\rm in}_{\rm max}\to \P _{\rm max}) = \omega_1 \omega_2 \frac{\P _{\rm max} - \P ^{\rm in}_{\rm max}}{2 \lambda \P _{\rm max}^2} .
\ee
Its maximal value, achieved for $\P _{\rm max} = 2\P _{\rm max}^{\rm in}$, is the Lyapunov stability time:
\be 
\tau_{0}(\P _{\rm max}^{\rm in})  = \frac{\omega_1\omega_2}{ 8 \lambda \P _{\rm max}^{\rm in}} .
\ee

\subsubsection{Stability at generic order}
The above discussion is easily generalized at order $n$. The residual time evolution is bounded by
\be 
\max_{i,\P '_i \leq \P _{\rm max}}\left| \frac{\partial}{\partial \Q_i} \delta H \right| 
 \leq 2 \epsilon^{n+1}  \max_{i,\P '_i \leq \P _{\rm max}} \sum_{N_1,N_2} \left| N_i (N_2 \omega_2 - N_1 \omega_1)  W_{\vec N}^{(n+1)} \right| \equiv \epsilon^{n+1} \P _{\rm max}^{n+2} \beta_n
\ee
where we included the factor of 2 due to higher orders, 
maximised over the free index $i=1,2$,
and used the fact that the remainder is a homogeneous polynomial in $\P '_i$ of order $n+2$. 
The function $\beta_n(\omega_1,\omega_2)$ can be computed numerically and diverges close to resonances:
\be 
\tau_{n}(\P ^{\rm in}_{\rm max}\to \P _{\rm max}) = \frac{\P _{\rm max} -\P _{\rm max}^{\rm in}}{ \epsilon^{n+1} \P _{\rm max}^{n+2} \beta_n } .
\ee
The Lyapunov stability time is
\be
\tau_{n}(\P _{\rm max}^{\rm in})  =  \frac{1}{\beta_n}\frac{(n+1)^{n+1}}{ (n+2)^{n+2}} \left( \frac{\omega_1\omega_2}{2 \lambda \P _{\rm max}^{\rm in}}\right)^{n+1}.
\ee
In view of the asymptotic character of the Birkhoff series, for each value of $\rho_0$, there is an optimal order $n$ that gives the strongest bound. 

\smallskip

As an example, in fig.\fig{StabilityBound}a we show the stability bound computed for $\omega_2/\omega_1 = \sqrt{2}$. 
The  numbers on the curve indicate the optimal order.
Some order dominates for a larger range when it contains enhanced
denominators. Our example contains enhanced denominators at 7th order ($1/(7 \omega_1-5\omega_2)$)
and 17th order.
Something similar happens  in fig.\fig{StabilityBound}b, 
where we consider $\omega_2/\omega_1 = \pi^2$. 
In both cases we keep fixed $\P'_{1,2}=1$ at any given order,
which approximatively means $\P_{1,2}=1$ for values of $\lambda$ small enough that the series converges.

For small enough coupling we proved ghost meta-stability up to cosmological times that cannot be probed
by numerical studies.  We consider a specific model that contains no special features: a similar analysis
can be performed for any other model.


%

\subsection{Resonances i.e.\ on-shell processes}\label{classres}
The previous perturbative approximation becomes less accurate close to resonances.
The most dangerous resonance corresponds to $\omega_1=\omega_2$,
as $1/(\omega_1-\omega_2)$ enhancements occur at leading order in the coupling.
As a result the Birkhoff series already fails for
$E_{\rm int}/E_{\rm free} \circa{>} (\omega_1-\omega_2)/\omega_{1,2}$,
instead of holding, as usual, when the energy in the interaction terms is smaller than the free energy.
Numerical solutions in our model with resonant $\omega_1\approx \omega_2$ (and very small $\lambda$ such that interactions
negligibly modify frequencies) show that a linear combination of $\P _{1,2}$ fails to
be quasi-constant of motion, but remains bounded so that run-aways remain avoided.

We extend analytic techniques to study resonances as they will be important in our subsequent study of 
classical and quantum field theories.\footnote{By expanding fields into Fourier modes one gets an infinite number of interactions,
that always contain resonances $\sum_i \omega_i^{\rm in} = \sum _j \omega_j^{\rm out}$
giving rise to decays and other on-shell process, using the standard terminology of quantum field theory
(when $E = \hbar \omega$ the resonance condition becomes conservation of energy and momentum).}


As described in advanced books about analytic mechanics~\cite{Arnoldone},
resonant processes can be analytically studied by modifying the Birkhoff normal form 
into a ``\emph{resonant} normal form'' that avoids the enhanced terms by selectively downgrading the goal
of  cancelling all dependence on the angle variables. 
One needs to keep those that give resonant combinations, obtaining a more complex but still manageable 
partially-diagonalised Hamiltonian.
Some combinations of $\P '$ remain quasi-conserved, whereas others evolve as governed by the resonant form.


\begin{figure}
$$\includegraphics[width=\textwidth]{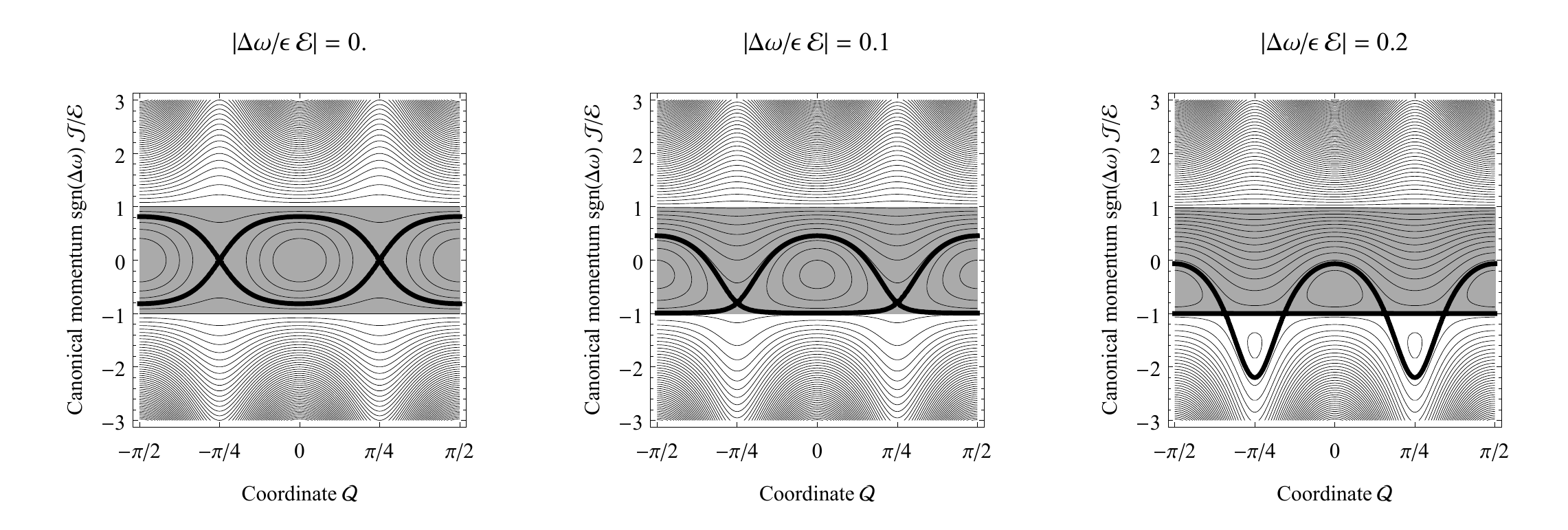}$$
\caption{\em Phase portrait of the auxiliary system close to the resonance.
The thick line is the separatrix between the different kinds of motion. The shaded gray region in phase space cannot be accessed with $\P '_{1,2} \geq 0$. 
 \label{fig:phaseportrait}}
\end{figure}

\subsubsection{Example: ghost that remains stable close to resonance}
To clarify with a worked example, we reconsider our model of eq.\eq{Hq1q2} in the resonant case 
$\omega_2 \to \omega_1 $.
We perform a  canonical transformation analogous to eq.\eq{W1model} (at leading order)
but omitting the singular Fourier modes with $N_1=N_2\equiv  \bar{N}$, that
multiply $\Q_1+\Q_2$.  
A straightforward but tedious change of variables gives
\be \label{eq:H'resmodel}
H' = \omega_1 \P '_1 - \omega_2 \P '_2 + \epsilon \frac{\P _1' \P _2'}{4} \left[ 1 + \frac{1}{2} \cos 2 (\Q'_1 + \Q'_2)\right]+\cdots.
\ee
The same result can  be re-obtained by expanding eq.\eq{H1model} in Fourier modes
and taking into account that off-diagonal elements of $W^{(1)}_{N_1 N_2}$ cancel the contribution from
$f_{N_1 N_2}$, while diagonal elements $W^{(1)}_{\bar{N}\bar{N}}$ vanish,
leaving the Hamiltonian Fourier coefficients $H^{(1)}_{\bar{N}} = \P _1' \P _2' f_{ \bar{N} \bar{N}} $ so that
\beq H^{(1)} = \sum_{ \bar{N}} e^{i  \bar{N} (\Q_1+\Q_2)} H^{(1)}_{ \bar{N}}
\ee
gives again eq.\eq{H'resmodel}, after taking into account that $\Q_i \simeq \Q'_i$.
The series expansion is no longer singular at the resonance, so that its first order is accurate at small coupling.
We can use it to study the dynamics close to the resonance finding that,
since $1  + \frac12 \cos 2 (\Q_1 + \Q_2)>0$, motion remains bounded.
This can be better seen by performing the canonical transformation 
\be 
\mathcal{Q} \equiv  \frac{\Q'_1+\Q'_2}{2} , \qquad  \mathcal{\P } \equiv \P '_1+\P '_2 , \qquad \mathcal{E} \equiv \P '_1 - \P '_2
\ee
such that, writing $\omega \equiv (\omega_1+\omega_2)/2$, $\Delta \omega \equiv \omega_1-\omega_2$, the Hamiltonian becomes
\be 
H' \simeq \omega \mathcal{E} + \Delta \omega \,
\frac{\mathcal{\P } }{2}+\frac{\epsilon}{16} (\mathcal{\P }^2 - \mathcal{E}^2)  \(1 + \frac12 \cos 4 \mathcal{Q}\).
\ee
$H'$ and $ \mathcal{E}$ are constants of motion,\footnote{In terms of original variables
the  `resonant'  constant of motion  $\mathcal{E} \equiv \P '_1 - \P '_2$ is
\beq \mathcal{E}=\P _1-\P _2+ \frac{\lambda \P _1 \P _2}{2\omega_1 \omega_2}\left[\frac{\cos2(\Q_1-\Q_2)}{\omega_1+\omega_2}
-\frac{\cos 2 \Q_1}{\omega_1}-\frac{\cos2\Q_2}{\omega_2}
\right]+{\cal O}(\lambda^2).\eeq
} 
while
$\mathcal{\P }$ is no longer conserved and forms, together with $\mathcal{Q}$,
a system with 1 degree of freedom, simple enough that can be analytically  studied.  
The key point is that its Hamiltonian is bounded so that $\mathcal{\P }$, despite not constant, is bounded and
the action variables $\P '_{1,2}$ are bounded too.
The possible motions are shown in fig.~\ref{fig:phaseportrait}. Typical trajectories move away from the resonance and then go back to it. 
${\cal \P }_{\rm max}/{\cal \P }_{\rm min}$ is generically of order one, with the
maximal variation $\sqrt{3}$ obtained for $\Delta\omega={\cal E}=0$. 
For $\Delta \omega$ sufficiently large, some of the trajectories in phase space oscillate. 
All trajectories are bounded.

\smallskip

In conclusion, the ghost system with quartic interaction $q_1^2 q_2^2$ is stable 
when perturbed around the non-interacting equilibrium point.
Away from resonances stability follows from the Birkhoff expansion and the KAM theorem~\cite{KAM,Arnoldone}; the latter states that away from resonances most trajectories in phase-space are still confined to be toroidal, even in the presence of small interactions. Close to the  $\omega_1 \simeq \omega_2$ resonance, stability follows because the extra
system is not a ghost, so its motion is bounded; 
higher-order resonances are not dangerous because their
resonant normal forms remain dominated by leading-order non-resonant terms.

\subsubsection{Example: ghost that undergoes run-away close to resonance}
The `safe' situation found in the previous model is not  generic. 
In other models a ghost can become unstable close to resonances.
This happens when the auxiliary dynamics that approximates the system close to a resonance 
is ghost-like and the resonant surface in phase space extending to 
$\P ' \to \infty$ (at fixed energy/approximate integrals of motion) is attractive. 

This happens, for example, replacing the quartic interaction $q_1^2 q_2^2$
with a cubic interaction $q_1^2 q_2$. The Hamiltonian in action-angle variables is
\be 
H = \omega_1 \P _1 - \omega_2 \P _2 + \epsilon \, \P _1 \sqrt{\P _2} \, \sin^2 \Q_1 \sin \Q_2
\ee
and the dangerous resonance is $\omega_2 \approx 2 \omega_1$ that (loosely speaking)
allows for a  $q_1 \to q_1 + q_2$ `decay'.
The resonant Birkhoff form at first order is
\be 
H' = \omega_1 \P _1' - \omega_2 \P _2' - \frac{\epsilon}{4} \P' _1 \sqrt{\P' _2} \sin (2 \Q'_1 + \Q'_2).
\ee
The sign of $\sin (2 \Q'_1 + \Q'_2)$ now qualitatively impacts the system.
This can be seen performing the 
canonical transformation $\mathcal{E} \equiv \P _1'-2 \P _2'$, $\mathcal{\P } \equiv (\P _1'+2\P _2')/4$, $\mathcal{Q}\equiv 2 \Q_1'+\Q_2'$ such that
\be 
H' = \tilde \omega \mathcal{E} + \Delta \omega \mathcal{\P } - \frac{\epsilon}{4} \( \frac{\mathcal{E}}{2} + 2 \mathcal{\P } \) \sqrt{\mathcal{\P } - \frac{\mathcal{E}}{4}} \, \sin \mathcal{Q}
\ee
with $\tilde \omega = (2 \omega_1 + \omega_2)/4$, $\Delta \omega = 2 \omega_1 - \omega_2$. The auxiliary system is now a ghost: the resonant ($\Delta \omega=0$) trajectories at fixed $\mathcal{E}$
extends to $\mathcal{\P } \to \infty$, e.g.\ the trajectory with $\mathcal{Q} = 0$. 
Moreover, these trajectories are attractive. 
At the resonance all trajectories are unbounded. Moving away from the resonance some stable KAM tori appear ``on one side'' for $\mathcal{\P }$ small enough, but nothing protects stability on the other side (large $\mathcal{\P }$). 



Notice that the condition of ghost safety is independent from the condition of bounded-from-below potential.
For example, consider a model with quartic interactions
$ H \supset \lambda  \,( q_1^2 q_2^2 + \kappa q_1^3 q_2)/2$.
Close to the resonance $\omega_2 \simeq 3 \omega_1$ we find that the ghost is safe for $|\kappa| < 2/\sqrt{3}$, despite the potential is unstable for any $\kappa \neq 0$ (for instance along the line $q_2=1$, $q_1 \to - \infty$). 
Conversely, the potential with with quartic interactions
$
H \supset \lambda'  \,( q_1^4 + \kappa q_1^3 q_2) $ 
is stable 
for any finite value of $\kappa$, but the ghost causes run-away for $|\kappa|>3 \sqrt{3}$.


The above considerations generalize to systems with more degrees of freedom. For instance, let us consider a system of 3 degrees of
freedom with interaction $q_1 q_2 q_3$, where $q_2$ is a ghost. The Hamiltonian in action-angle variables is
\be 
H = \omega_1 \P _1 - \omega_2 \P _2 + \omega_3 \P _3 + \epsilon \sqrt{\P _1 \P _2 \P _3} \sin \Q_1 \sin \Q_2 \sin \Q_3.
\ee
The first-order resonant form close to the dangerous resonance  $\omega_1 - \omega_2 + \omega_3 \equiv \Delta \omega \simeq 0$ is
\be 
H \simeq  - \omega_2 \mathcal{E}_2 + \omega_3 \mathcal{E}_3 + \Delta \omega \frac{\mathcal{\P }}{3} - \frac{\epsilon}{4} \sqrt{\frac{\mathcal{\P }}{3}\( \frac{\mathcal{\P }}{3} + \mathcal{E}_2\)\( \frac{\mathcal{\P }}{3} + \mathcal{E}_3\)} \, \sin 3 \mathcal{Q} \,
\ee
where $\mathcal{E}_i \equiv \P _i'-\P _1'$,  $\mathcal{Q} = (\Q'_1+\Q'_2+\Q'_3)/3$ and $\mathcal{\P } = 3 \P _1'$. The extra-system Hamiltonian is unbounded and as a consequence 
the system, on resonance, undergoes ghost run-away.

\smallskip

The discussion of various examples allows to identify a useful general property:
only the part of the Hamiltonian at most quadratic in $\P'$ is typically relevant for stability, since close enough to the origin cubic and quartic interactions dominate over higher orders. 
In the presence of both cubics and quartics, 
 quartic interactions generically stabilise the otherwise un-safe behaviour of cubic-only interactions. 
This can be seen by noticing that  resonant normal forms of quartic interactions contain stabilising terms $\sim \P'^2$ (as in eq.\eq{H'resmodel}), that dominate  with respect to the dangerous dynamical terms $\sim \P'^{3/2} f(\Q)$ for sufficiently large $\P$. 

In conclusion, ghost stability in classical mechanics is generic at small coupling away from resonances.
In most models, resonances do not lead to ghost run-away but only to partial energy flow.

\section{Ghost meta-stability in quantum mechanics?}\label{QM}
Moving from classical to quantum mechanics, we again consider the prototype model of eq.\eq{Ltoyq},
described by the Hamiltonian
\beq \label{eq:Hq1q2bis} H = \frac{p_1^2}{2} - \frac{p_2^2}{2} + V,\qquad
V = \omega_1^2 \frac{q_1^2}{2}- \omega_2^2 \frac{q_2^2}{2}+\frac{\lambda}{2} q_1^2 q_2^2\eeq
which leads to the Schroedinger equation for the wave-function $\psi(q_1, q_2)$
\beq \label{eq:Schrghost}
- \frac{\hbar^2}{2}  \frac{\partial^2 \psi}{\partial q_1^2} +
\frac{\hbar^2}{2}  \frac{\partial^2 \psi}{\partial q_2^2} = (E-V) \psi.
\eeq
We remind the following features of the Schroedinger equation
in the absence of ghosts and relevant for computing vacuum tunnelling through a potential barrier:
1) the sign of $E-V$ tells in which regions $\psi$ oscillates or gets exponentially suppressed;
2) the vanishing of $E-V$ determines the  `release point' $q_*$ on the other side of the potential barrier
after which classical motion is unstable;
3) the tunnelling rate is exponentially suppressed by the WKB bounce action $W = \min \int_0^{q_*} dq \sqrt{2V}/\hbar$,
where the integral is along the path in multi-dimensional field space that minimises $W$.

These features are now lost because the ghost appears with an opposite sign in eq.\eq{Schrghost}.
So the classically meta-stable ghost $q_2$ might become unstable if
the wave-function $\psi(q_1, q_2)$ of any state
extends along the  classically-allowed region
$q_1 \approx q_2$ reaching the large values where classical motion leads to run-away.

\subsection{Model computation}
In the presence of a ghost an infinite numbers of states have $E =0$, or any other value. 
The same happens, without ghosts, in the presence of a potential like $V = \omega^2 q^2/2 + \lambda q^4/2$
with negative $\lambda$: despite that $V$ is unbounded-from-below, the lowest-energy bound state is special.
We focus on the analogous of this state for the ghost system.
In the free theory such ground-like bound state has minimal positive energy and maximal negative energy.
Thanks to this property, it might be selected by cosmological evolution.
We now show that the ground-like state is meta-stable.

 \begin{figure}[t]
$$\includegraphics[width=0.35\textwidth]{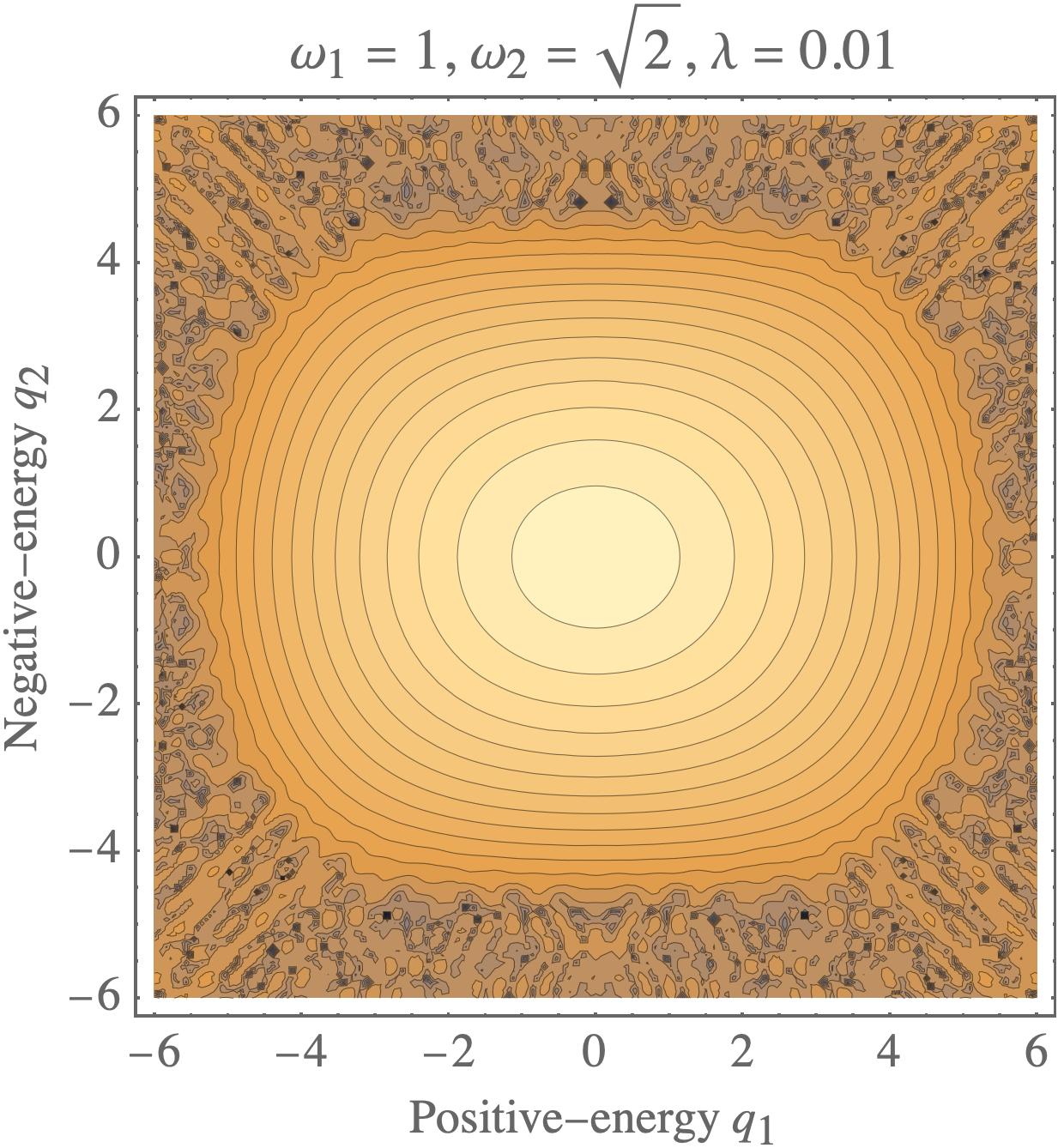}\qquad
\includegraphics[width=0.35\textwidth]{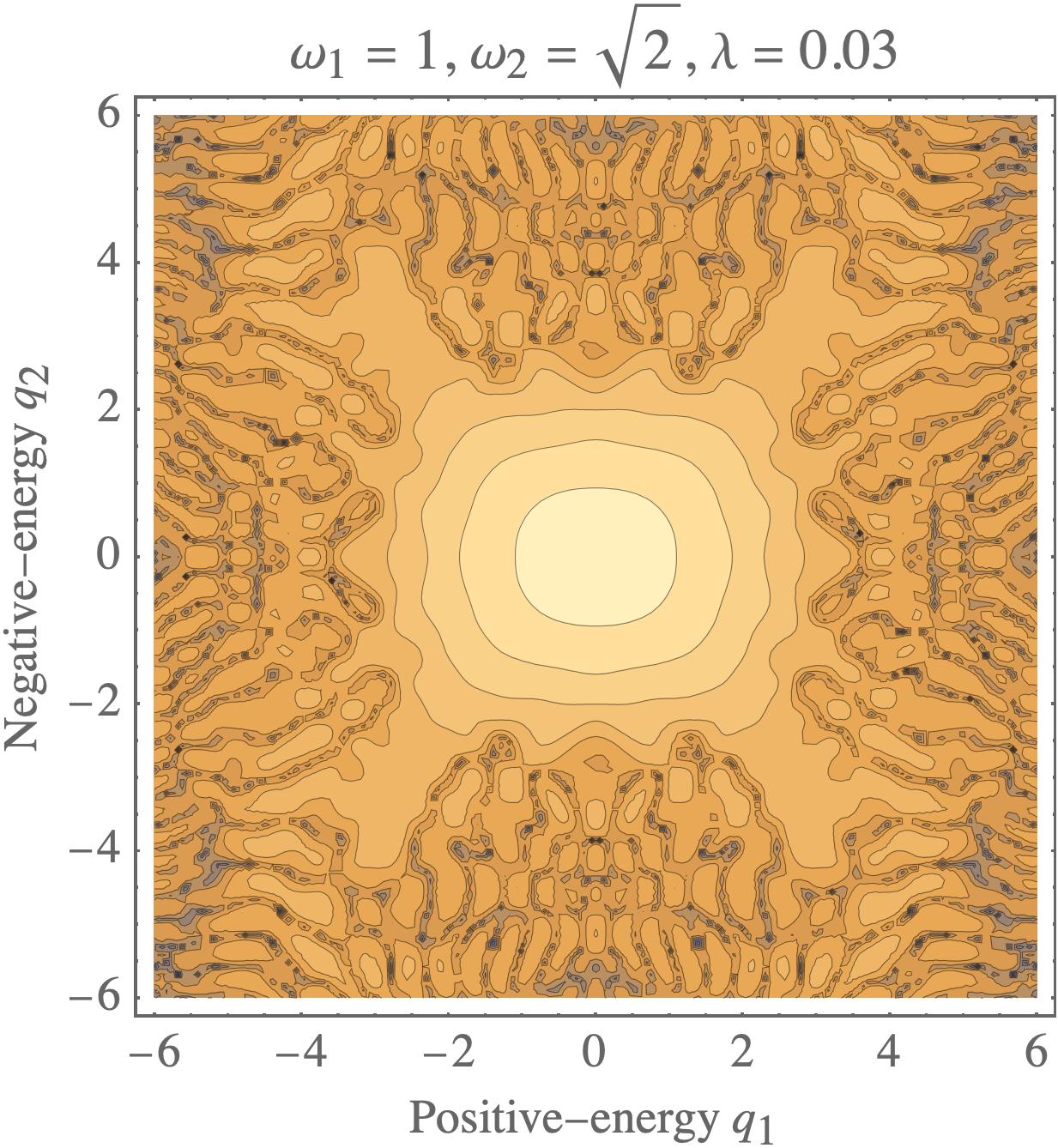}
$$
\caption{\em Iso-curves of the ground-like state wave-function $|\psi(q_1,q_2)|^2$ for different
values of the quartic coupling $\lambda$ between the positive-energy $q_1$ and the negative-energy $q_2$. 
Contour curves are separated by one order of magnitude.
\label{fig:psi}}
\end{figure}

We start by numerically computing the ghost model described by the Hamiltonian of eq.\eq{Hq1q2bis}.
If the coupling $\lambda$ vanishes it reduces to two decoupled harmonic oscillators, with the usual eigenstates
$\ket{n_1, n_2}$.
The ground-like state is $\ket{0,0}$ with wave-function $\psi_{00} (q_1,q_2) = \psi_0(q_1) \psi_0(q_2)$
with $\psi_0(q_i) \propto e^{-q_i^2 \omega_i/2\hbar } $.
For $\lambda\neq 0$ the ground-like state is  the one that tends to $\ket{0,0}$ as $\lambda\to 0$,
and that thereby, at small $\lambda$, has maximal projection along $\ket{0,0}$.
Its wave function $\psi(q_1,q_2)$ has no nodes around $q_1\sim q_2\sim 0$ and
can be computed  either numerically solving the Schroedinger eq.\eq{Schrghost}
%
or by writing the Hamiltonian $H $ of eq.\eq{Hq1q2bis} as a matrix in the
$\ket{n_1, n_2}$ basis and diagonalising it.
Matrix elements of the interaction term $\lambda q_1^2 q_2^2/2$ are  computed using
\beq \md{q^2_i}_{n_im_i}=\frac{\hbar}{2\omega_i}
\left\{\begin{array}{ll}
 \sqrt{(m_i+1)(m_i+2)} & n_i = m_i+2\cr
 \sqrt{(n_i+1)(n_i+2)} & m_i = n_i+2\cr
2n_i+1 & n_i = m_i\cr
0 &\hbox{otherwise .}
\end{array}\right.\eeq
Fig.\fig{psi} shows examples of numerical results in a non-resonant case $\omega_1\neq \omega_2$:
the ghost model gives a $|\psi(q_1, q_2)|^2$ 
qualitatively similar to what obtained in a model with two positive-energy $q_{1,2}$
and an unbounded-from-below potential with $\lambda <0$.
Inside the barrier at $q_1\sim q_2 \sim 0$ the wave-function is the usual Gaussian;
outside it has an oscillatory pattern with exponentially suppressed amplitude.
In our approximation the wave-function is real, but one can compute a more accurate
bound-state with complex wave-function such that the
exponentially suppressed probability current is out-flowing  only.
Its flux equals the vacuum decay rate, and the
energy eigenvalue acquires a correspondingly exponentially suppressed imaginary part (see e.g.~\cite{Bender:1990pd}).

The ghost case qualitatively differs from the negative-potential case only in the
resonant situation $\omega_1 = \omega_2$:
the ghost ground-like state does not reduce to $\ket{0,0}$
as $\lambda\to 0$. 


\subsection{The WKB approximation}
Ghost meta-stability can be understood more in general taking into account
that  tunnelling can be approximated {\em a la} WKB.
Writing the wave function as $\psi = e^{iS/\hbar}$, the Schroedinger equation reduces to the classical Hamilton-Jacobi (HJ) equation
\beq \label{eq:HJ}\frac{\partial S}{\partial t} = - H \bigg(q_i, p_i=\frac{\partial S}{\partial q_i}\bigg)\eeq
plus extra terms $\frac12 i\hbar \partial^2 S/\partial q_i^2$ 
neglected at leading order in the semi-classical expansion,
which is enough to approximate vacuum decay at weak coupling.

In Hamiltonian mechanics, eq.\eq{HJ} is obtained by demanding that
$S$ generates a classical canonical transformation such that the transformed Hamiltonian vanishes.
Its solution is the  classical action $S(q,t) = \int^{q,t}_{0,0} L(q_{\rm cl})\,dt$
computed along the classical particle trajectory going from $q=0$ at time $t=0$ to $q$ at time $t$.
Thereby the HJ wave equation provides a bridge between waves and particles:
$S$ respects the good hidden properties of a classical ghost discussed in section~\ref{class}.
To make better contact with the formalism of section~\ref{class} we 
consider a Hamiltonian $H$ that does not depend on time.
Then eq.\eq{HJ} can also be solved by separating variables as
$S(q,t) = W(q)-E t$ where $E=H$ is the constant energy and
$W$ generates a canonical transformation to action-angle variables $(\Theta_i ,J_i)$
such that $H$ only depends on $J_i$.
The `reduced action' $W$ satisfies the wave equation
\beq \label{eq:HJW} E = H(q_i, p_i =\frac{ \partial W}{\partial q_i})\qquad \Rightarrow\qquad W  =  \int\, p_i\,dq_i.\eeq
The classical change of variables to action-angle coordinates essentially is a `diagonalization' of the  classical Hamiltonian.
Eq.\eq{HJ} (eq.\eq{HJW}) approximates the time-dependent (time independent) Schroedinger equation eq.\eq{Schrghost},
with the first (second) form being more useful for computing the propagator (energy eigenstates).

The hidden constants of motion that in the classical theory
forbid  motion into the dangerous region $q_1 \approx q_2$
still play a role in the semi-classical approximation.
No new dramatically fast ghost instabilities appear in the quantum theory as,
going away from the origin $q_1\sim q_2 \sim 0$,
the wave function gets exponentially suppressed by the semi-classical WKB factor $W$.
Having a quantum Hamiltonian in action-angle variables, $H = \omega(\P ) \P $,
its eigenstates are the $\ket{\P }$ states with eigenvalues $E=H(\P )$
and wave function $\langle  \Q| \P  \rangle = e^{i \P \Q/\hbar}$,
so that its periodicity demands $\P =n \hbar$ with $n$ an integer.




\medskip

To obtain tunnelling rates we need to compute how the 
wave-function extends into the classically forbidden region:
as well-known it is useful to perform an analytic continuation to
Euclidean time, $t_{\rm E} = i t$ and solve the Euclidean HJ equation
with $L_{\rm E} = \frac12 (d\vec q/dt_{\rm E})^2 - V_{\rm E}$ and inverted potential $V_{\rm E} = - V$.
A well-known computational simplification allows to approximate
potential tunnelling in the absence of ghosts:
the vacuum decay rate is approximated by $e^{-B}$, where
the bounce action $B=\min W_{\rm E}$ is computed along the
classical Euclidean trajectory in field space that connects the false vacuum to the
other side of the potential barrier with minimal $W_{\rm E}$.
For example
\beq \label{eq:WKB+}
B= \min W_{\rm E} = \min S_{\rm E} = \min \lim_{t_{\rm E}\to +\infty} \int_{0,0}^{\vec q_*, t_{\rm E}} L_{\rm E}\, dt_{\rm E} =\min \int_0^{\vec q_*} dq \sqrt{2V_{\rm E}}\eeq
for the ground state with $E\to 0^+$.
This simplification holds in the presence of multiple degrees of freedom, and thereby allows to compute 
vacuum decay in Quantum Field Theory~\cite{Coleman:1977py}.

A similar result holds in the presence of ghosts only, 
with the only difference that boundary conditions (normalizable wave-function) now demand 
picking the opposite-sign solution to the HJ equation.
The sign of $W$ is not fixed because $H$ contains $p^2 =(\partial W/\partial q)^2$.
For the ground state $E\to 0^-$ the bounce action is similar to eq.\eq{WKB+}
but with $t_{\rm E}\to - \infty$.
Equivalently, an opposite-sign Wick rotation is needed to make the Euclidean ghost action positive.

In the presence of positive-energy particles that interact with ghosts,
the desired solution to the HJ equation
can be found numerically or perturbatively up to $q^2 \circa{<}\omega/\lambda$, 
\beq W_{\rm E}(q_1,q_2)|_{E=0} =\frac{1}{2} q_1^2 \omega _1+\frac{1}{2}
   q_2^2 \omega _2+\frac{\lambda  q_1^2
   q_2^2}{4 \left(\omega _1-\omega _2\right)}+
\frac{\lambda ^2 \left(q_2^2 q_1^4 \omega _1-2 q_2^2 q_1^4 \omega _2-2 q_2^4 q_1^2 \omega
   _1+q_2^4 q_1^2 \omega _2\right)}{16  \left(\omega
   _1-\omega _2\right){}^2\left(2 \omega _1-\omega _2\right)\left(\omega _1-2 \omega _2\right) }+\cdots\eeq
but we don't know how to compute vacuum decay bypassing a full solution to the HJ equation~\cite{Knudson}.
Physically, the new complication arises because we are interested in the ground-like state,
which is neither the lowest nor the highest energy state, so that selecting it gets more complicated.

\section{Ghost meta-stability in classical field theory?}\label{CFT}
A field $\varphi(\vec x,t)$ can be decomposed as an infinite number of Fourier modes $q_{\vec n}(t)$.
An infinite numbers of degrees of freedom allows for new phenomena.
Some of them make any  interacting classical field theory problematic, 
others are a problem for theories containing ghosts.
As ghosts are at most a co-morbidity of the theory,
one needs to address and disentangle the new intertwined issues.
\begin{enumerate}
\item
In order to compute numerically one has to `regularise' the theory by introducing 
a cut-off on the number of degrees of freedom,
usually realised by a minimal length $a$, such as a lattice discretisation of space-time.
Typical discretised field equations do no conserve energy and can
lead to fake run-away behaviours when evolving configurations with excited
modes near the cut-off (the ones where energy conservation is badly violated).
We will  define special discretised classical
equations that exactly conserve total energy,
but hidden pseudo-constants of motion can be violated by the regularisation.

\item At some moment and in some region of space,
some modes can acquire a higher energy density 
and overcome the energy barrier between stability and instability.
In thermal field theories with local minima in the potential
this is the well-known thermal tunnelling, characterised by a 
space-time tunnelling probability density.\footnote{Some authors claim that they can approximate quantum
vacuum decay rate by classically evolving a field starting from quantum-like
initial conditions~\cite{hep-th/9110037,1806.06069} and waiting for a large enough energy fluctuation that
goes over the potential barrier.
However this can only be a rough approximation, because
an interacting classical field theory tends to evolve towards a thermal state
where energy is equipartitioned among all modes.}
The same mechanism contributes to ghost instabilities.

\item General initial field configurations tend to thermalise.
However, a thermal state is impossible in classical field theory, as each one of the infinite modes
should have the same energy $\sim T$.
In electro-magnetism, this is the well known black-body problem.
An interacting field theory gives rise to a cascade of energy towards higher-frequency modes,
and the temperature evolves towards $T\to 0$. 
On a lattice, this cascade stops when the problematic modes at the cut-off thermalise.

\item The above issue is solved by quantum mechanics.
For a thermal state, classical field theory only holds for modes with $E \circa{<} T$, 
and is replaced by quantum field theory for modes with $E\circa{>} T$
that get suppressed energy density:
\beq f = \frac{1}{e^{E/T}-1 },\qquad  f E \simeq \left\{\begin{array}{ll}
T & E\ll T\cr
E\, e^{-E/T} & E \gg T
\end{array}\right.  .\eeq 

\item Finally, the main new point.
Field theory contains an infinite number of modes $q_{\vec n}(t)$ with  frequencies $\omega_n$, 
so resonances  are always possible.  
These resonances are the usual on-shell processes
such as decays and scatterings.  
In the presence of  ghosts, resonances can lead to partial or total loss of hidden constants of motion
as discussed in section~\ref{classres}.
\end{enumerate}
In section~\ref{phi_k} we decompose fields $\varphi(x,t)$ into modes $q_n(t)$,
and in section~\ref{fieldKAM} we perform a stability analysis of the resonances:
hidden constant of motion persist up to ${\cal O}(1)$, but the number of resonance is so large
that dangerous  energy transfer between normal fields and ghosts can take place.
As a consequence, assuming no protection, in section~\ref{thermalizationrate} we use
statistical methods to compute the energy transfer between normal fields and ghost fields.
Finally, in section~\ref{NumericalRes} we compare analytic results to
numerical classical lattice simulations (using the convenient discretised field equations 
described in appendix~\ref{lattice}).


\subsection{Classical equations of motion in momentum space}\label{phi_k}
We consider a scalar field $\varphi(x,t)$ in 1+1 dimensions.
In a box $0\le x \le L$ with periodic boundary condition the scalar field is expanded in normal modes $q_n$ as
\beq \varphi(x,t) = \frac{1}{\sqrt{L}}\sum_{n=-\infty}^\infty   q_n(t) e^{i k_n x} \qquad
k_n = \frac{2\pi n}{L} .
\eeq
We consider a real scalar field, so $q_{-n} = q_n^*$.
The Lagrangian density $\Lag_\varphi= (\partial_\mu\varphi)^2/2 - m^2 \varphi^2/2 + \Lag_I$ gives
the Lagrangian
\beq {\cal L} = \int_0^L dx \,\Lag_\varphi = 
\frac{\dot q_0^2 - m^2 q_0^2}{2}+
\sum_{n=1}^\infty ( |\dot q_{n}|^2 - \omega_n^2 |q_n|^2) + {\cal L}_I,\qquad \omega_n^2 = m^2 + k_n^2.
\eeq
The $dx$ integral is simply given by $L$ times the expansion of $\Lag$, keeping only those terms such that
their $e^{ikx}$ factors multiply to 1.
The classical equations of motion are
\beq \ddot q_n  + \omega_n^2 q_n =  \frac{\partial {\cal L}_I}{\partial q_n}.\eeq
Classical evolution can be restricted to real $q_n$, which means zero momentum for each mode.
The averaged free classical Hamiltonian is
\beq \langle H \rangle =  \int dx \frac{1}{2} \langle\dot \varphi^2+\varphi^{\prime 2} + m^2 \varphi^2 \rangle = \sum_{n=-\infty}^{+\infty} \omega_n^2 \langle q_n q_{-n} \rangle \eeq
so that the classical thermal state with equipartition of relativistic energy corresponds to
$q_n = \sqrt{T}/\omega_n$, which is the (UV divergent) classical limit of the Bose-Einstein distribution,
$\med{{q}_n {q}_{-n}} =  \hbar (1/2 +f_n)/\omega_n$ 
with $f =1/(e^{E/T}-1) \to T/E \gg1$ at $E\ll T$.
The extra $1/2$ is the purely quantum fluctuation.
$\langle H \rangle$ is UV divergent both in classical physics at finite temperature $T$, and in quantum physics.

\subsection{Analytic study of one ghost resonance in field theory}\label{fieldKAM}
As a prototypical field theory containing a normal field $\varphi_1$ interacting with a ghost field $\varphi_2$
we consider the Lagrangian of eq.\eq{Lagphi12}
where the ghost is obtained setting $\pm = -1$.
For simplicity we here compute in 1+1 dimensions, as this is enough to encounter the new key phenomena.
The two fields $\varphi_{1,2}$ have positive and negative kinetic energy, respectively.
We expand each of them in normal modes $q_{n_1}$ and $q_{n_2}$ as outlined in the previous section.
The interactions among momentum modes $q_{n_i}$ are complicated because locality is not manifest.
Let us focus on four generic modes:
$n_1$ and $n'_1$ for $\varphi_1$ and
$n_2$ and $n'_2$ for $\varphi_2$.
We assume that $k_{n_1}+k_{n'_1}+k_{n_2}+k_{n'_2}=0$.
Then, their interaction term is
\begin{align}
 \int d x \, \varphi_1^2 \varphi_2^2 &= \frac{4}{L} (q_{n_1} q_{n'_1}  q_{n_2} q_{n'_2} + 
 q_{-n_1} q_{-n'_1}  q_{-n_2} q_{-n'_2} +
q_{n_1}q_{-n_1}q_{n_2}q_{-n_2} +   \notag\\
&  +
q_{n'_1}q_{-n'_1}q_{n_2}q_{-n_2} +
q_{n_1}q_{-n_1}q_{n'_2}q_{-n'_2} +
q_{n'_1}q_{-n'_1}q_{n'_2}q_{-n'_2} +
\cdots).
\end{align}
The frequencies are generically off-resonance but  for some choice of momenta
they satisfy resonant conditions 
such as $N_1 \omega_{n_1}  +N_2 \omega_{n'_1} -N_3  \omega_{n_2} - N_4\omega_{n'_2}$
even for $N_i = \pm 1$, giving rise to on-shell processes. 


We isolate a sub-system of four such degrees of freedom $q_{n_i}$.
For simplicity we can assume that their initial conditions are real, so that they remain real and we can
treat $q_n = q_{-n}$ as a single degree of freedom.  
Moving to action-angle variables and simplifying the notation,
we write their pulsations as $\omega_{1,2,3,4}$ and their
actions as  $\P _{1,2}$ (positive energy)
and $\P _{3,4}$ (negative energy).
The Hamiltonian of the sub-system is
\begin{align}
H = \, &\omega_1 \P _1 + \omega_2 \P _2 - \omega_3 \P _3 - \omega_4 \P _4+ \epsilon \, \bigg(  \frac{\P _1}{\omega_1} \frac{\P _3}{\omega_3} \sin^2 \Q_1 \sin^2 \Q_3 +  \frac{\P _1}{\omega_1} \frac{\P _4}{\omega_4} \sin^2 \Q_1 \sin^2 \Q_4  +\\
& +  \frac{\P _2}{\omega_2}\frac{\P _3}{\omega_3}\sin^2 \Q_2 \sin^2 \Q_3 + \frac{\P _2}{\omega_2}\frac{\P _4}{\omega_4} \sin^2 \Q_2 \sin^2 \Q_4+ 2 \sqrt{\frac{\P _1}{\omega_1}\frac{\P _2}{\omega_2}\frac{\P _3}{\omega_3}\frac{\P _4}{\omega_4}} \sin \Q_1 \sin \Q_2 \sin \Q_3 \sin \Q_4 \bigg) .\notag
\end{align}
where  $\epsilon = 8\lambda/L $.
Off-resonance the system is stable, and we now study the possibly dangerous resonant case,
assuming $\omega_1 + \omega_2 - \omega_3- \omega_4 \equiv \Delta \omega \simeq 0$.\footnote{We assume for now that no other combinations are vanishing, so that resonances do not ``overlap''. In appendix~\ref{sec:overlapping} we show that the case of all frequencies close to each other leads to similar conclusions as the ones discussed here.}
Close to resonance, the normal resonant form at leading order is
\begin{align}
H \simeq \, &\omega_1 \P _1' + \omega_2 \P _2' - \omega_3 \P _3' - \omega_4 \P _4'+ \frac{\epsilon}{4} \, \bigg(  \frac{\P _1'}{\omega_1} \frac{\P _3'}{\omega_3}  +  \frac{\P _1'}{\omega_1} \frac{\P _4'}{\omega_4}   +  \frac{\P _2'}{\omega_2}\frac{\P _3'}{\omega_3} + \frac{\P _2'}{\omega_2}\frac{\P _4'}{\omega_4} +\\
& + 2 \sqrt{\frac{\P _1'}{\omega_1}\frac{\P _2'}{\omega_2}\frac{\P _3'}{\omega_3}\frac{\P _4'}{\omega_4}} \cos (\Q_1'+\Q_2'+\Q_3'+\Q_4') \bigg) .\notag
\end{align}
We  isolate the auxiliary system by the canonical change of variables generated by 
\beq W=\mathcal{\P }(\Q'_1+\Q'_2+\Q_3'+\Q_4')/4 + \mathcal{E}_2 \Q_2'+ \mathcal{E}_3 \Q_3'+ \mathcal{E}_4 \Q_4'\eeq
i.e.\ $4{\cal Q} = \Q_1'+\Q_2'+\Q_3'+\Q_4'$ and $\P '_1={\cal \P }/4$, $\P '_i={\cal \P }/4+{\cal E}_i$. 
The resonant form becomes
\begin{align}
H &\simeq \omega_2 \mathcal{E}_2 - \omega_3 \mathcal{E}_3 - \omega_4 \mathcal{E}_4 + \Delta \omega \frac{\mathcal{\P }}{4} + \frac{\epsilon}{4} \bigg[ \frac{1}{\omega_1 \omega_3}\frac{\mathcal{\P }}{4}\(\frac{\mathcal{\P }}{4} + \mathcal{E}_3\) +  \frac{1}{\omega_1 \omega_4} \frac{\mathcal{\P }}{4}\(\frac{\mathcal{\P }}{4} + \mathcal{E}_4\) + \notag\\
&+  \frac{1}{\omega_2 \omega_3} \(\frac{\mathcal{\P }}{4} + \mathcal{E}_2\) \(\frac{\mathcal{\P }}{4} + \mathcal{E}_3\) +  \frac{1}{\omega_2 \omega_4}\(\frac{\mathcal{\P }}{4} + \mathcal{E}_2\) \(\frac{\mathcal{\P }}{4} + \mathcal{E}_4\) +\notag\\
& \frac{1}{\sqrt{\omega_1 \omega_2\omega_3 \omega_4}}\sqrt{\frac{\mathcal{\P }}{4}\(\frac{\mathcal{\P }}{4} + \mathcal{E}_2\)\(\frac{\mathcal{\P }}{4} + \mathcal{E}_3\)\(\frac{\mathcal{\P }}{4} + \mathcal{E}_4\)} \cos 4 \mathcal{Q} \bigg]
\end{align}
so that ${\cal E}_{1,2,3}$ are constant of motion i.e.\ all $\P '_i$ vary by a common amount ${\cal \P }/4$.
The important result is that $\cos 4 \mathcal{Q}$ cannot dominate over the sum of other terms,
so that this resonance does not lead to ghost run-away, but only to a partial violation
up to ${\cal O}(1)$ factors of the hidden conservation law.
This means that the local interaction $\varphi_1^2 \varphi_2^2$
of field theory gives, when expanded in normal modes,
a specific set of interactions among them such that each on-shell resonance 
allows an order one energy transfer among the modes, but no ghost run-away.



\subsection{Analytic study of multiple ghost resonances in field theory}\label{fieldKAM2}
We next need to study what is the collective effect of the infinite number of such resonances present in the
continuum limit: the number of modes $N = L/a$ diverges
when the lattice cut-off $a$ becomes infinitesimally small, or the box size $L$ infinitely large.
The Hamiltonian in action-angle variables is an infinite sum of terms like those discussed in the previous section
\begin{align}
H &= \sum_{n_1 = - \infty}^{+ \infty} \omega_{n_1 }\P _{n_1}-
\sum_{n_2 = - \infty}^{+ \infty} \omega_{n_2}\P _{n_2} \notag\\
&\;+ \; \epsilon  \!\!\!\!\!\!  \sum_{n_1, n'_1, n_2, n'_2} \!\!\!\!\!\! \delta_{0,n_1 + n'_1 + n_2 + n'_2} \sqrt{\P _{n_1} \P _{n'_1} \P _{n_2} \P _{n'_2}} \sin \Q_{n_1} \sin \Q_{n'_1} \sin \Q_{n_2} \sin \Q_{n'_2} ,
\end{align}
with $\epsilon = 2\lambda/L (\omega_{n_1} \omega_{n'_1} \omega_{n_2} \omega_{n_2})^{1/2}$. 
The rough argument goes as follows. 
At small coupling the theory contains $2N$ quasi-integral of motion: one for each degree of freedom.
In the continuum limit the number of resonances scales as $N^2$ 
(out of the 4 momenta, 2 combinations are fixed by momentum conservation and resonance condition, i.e.\ energy conservation). 
Each resonance produces the partial loss of a quasi-integral of motion $\mathcal{E}$. 
Asymptotically, all quasi-integrals of motion are lost and the available phase-space is filled up,
allowing for ghost run-away.

\medskip

The argument above can be made more precise. A combination is resonant if the detuning $\Delta \omega \equiv \omega_{n_1} + \omega_{n'_1} - \omega_{n_2} - \omega_{n_2}$
is smaller than the expansion parameter $\epsilon  \P  $, where $ \P  $ is the typical value of the actions, 
e.g.\ $ \P   = T/\omega$ for a thermal state. 
For finite $L$ the resonance is not exactly satisfied and the expansion parameter is finite. Both quantities go to zero in the continuum limit, so a careful analysis is needed. 
Let us consider modes up to an UV cut-off $k \lesssim k_{\rm max}$. 
A  resonance that would be perfect in the continuum acquires, in view of the discreteness $\delta k = 2 \pi/L$,
a typical detuning
$ \Delta \omega \approx  (8 \pi/L)  (k_{\rm max} /\omega_{\rm max})$.
Here $\omega_{\rm max}$ is the frequency corresponding to $k_{\rm max}$ having ignored, for simplicity, that it differs for fields $\varphi_1$ and $\varphi_2$ if $m_1\neq m_2$.
The fraction of such interactions that are resonant for finite $L$ is $f = \epsilon  \P / \Delta \omega$.
This stays finite in the continuum limit, as both $\epsilon$ and $\Delta \omega$ scale as $1/L$.
So the ghost is not protected when $f N^2 \gtrsim N$ i.e.\ $N =L/a \gtrsim1/f \sim \omega^3/\lambda T$.
Then the action $J_n$ of one typical microscopic mode can change by order one on a time-scale
$\Gamma \sim \lambda T/\omega^2$, linear in $\lambda$ at leading order.
As discussed in the next section, the macroscopic properties of the system evolve on a slower time-scale
$1/\tau=\Gamma/N \sim \lambda^2 T^2/\omega^5$. 
As we will see, this is the scale of the instability time. If, instead, there were no microscopic protection for the single modes, the instability time would have been much faster, linear in $\lambda$. 

\medskip


\subsection{The ghost run-away rate}\label{thermalizationrate}
Based on the previous discussion we assume that 
the extra quasi-conserved energies get violated  in field theory by resonances.
Then the system evolves statistically, towards the direction that increases total entropy $S=S_1+S_2$,
where 1 is the positive-energy sector and 2 is the ghost.
We define the ghost temperature $T_2$ as the average ghost energy $E_2\le 0$ per degree of freedom, $T_2 = E_2/N\le 0$. 
Let us compute $S_2$.
The volume in phase space is easily found in action-angle variables: 
\be
{\cal V}_2 = (2 \pi)^N \, \frac{N^N |T_2|^N}{N !} .
\ee
The factor of $(2\pi)^N$ is the contribution of the angle variables, whereas the remaining factor is the volume of the simplex $\sum \omega_n \P _n \leq |E_2|$. Therefore the ghost entropy is 
\be
S_2 = N \log |T_2|
\ee
up to a $T_2$-independent constant.
The total entropy $S=S_1+S_2$ of the system at fixed total energy $E_1+E_2$
is maximal when
\be 
\delta S = \frac{\partial S}{\partial E_1} \, \delta E_1 \;+\; \frac{\partial S}{\partial E_2} \, \delta E_2 = \delta E_1 \( \frac{1}{T_1} - \frac{1}{T_2}\) = 0 
\ee
which can only occur for $T_1 \to \infty$ and $T_2\to-\infty$.
Heat flows from the ghost to the positive-energy system and the thermodynamic evolution eventually causes the run-away on a time-scale $\tau$, that we now compute.


We consider a theory in $d$ spatial dimensions with the Lagrangian of eq.\eq{Lagphi12}. 
To set the formalism, we first assume that both fields $\varphi_{1,2}$ have positive kinetic energy.
Then, starting from temperatures $T_{1,2}\ge 0$
they thermalise towards the equilibrium state with a common temperature $T=(T_1+T_2)/2$
via the $\lambda \varphi_{1}^2 \varphi_2^2/2$ interaction.
The thermalization process can be computed using Boltzmann equations.
We consider their well known quantum expression and perform its classical limit,
to later compare with numerical classical evolution on a lattice.
In order to keep $\hbar$ factors explicit it is convenient to express quadri-momenta $P_\mu$ in terms
of wave vectors, $P_\mu = (E, \vec p) = \hbar K_\mu = \hbar (\omega, \vec k)$.
The Lagrangian $\Lag$ contains no $\hbar$ factors, so the mass parameters $m_{1,2}$ have dimension 1/time.
The contribution of $12\leftrightarrow  1'2'$ scatterings to the Boltzmann equation for the energy density $\rho_1$
(assumed to be spatially homogeneous) of $\varphi_1$ at leading order in the interaction $\lambda$ is
\beq \label{eq:drho1gen}
\dot \rho_1 = - \int d\vec k_1d\vec k_2d\vec k'_1d\vec k'_2\,
E_1\, (2\pi)^{d+1}\delta(K_1+K_2-K'_1-K'_2)  |\mathscr{A}|^2 F\eeq
where $\mathscr{A} =2 \hbar \lambda$ is the amplitude;
$d\vec{k} = d^dk/2\omega(2\pi)^3$ is the usual relativistic phase space;
one can symmetrise $E_1\to (E_1-E'_1)/2$.
Finally $F$ depends on particle number densities
$dn_i = f_i \,d^dk_i/(2\pi)^d$:
\beq F =   f_1(E'_1) f_2(E'_2)[1+  f_1(E_1)][1+ f_2(E_2)] 
-  f_1(E_1)   f_{2}(E_2) [1+ f_1(E'_1)][1+ f_2(E'_2)] .\eeq
It vanishes when Bose-Einstein distributions $f(E) =1/(e^{E/T}-1)$ realise thermal equilibrium.
Total energy is conserved, so $\dot\rho_2 = - \dot\rho_1$.
The quantum Boltzmann eq.\eq{drho1gen} has two classical limits: particle and wave.
The particle limit corresponds to small occupation numbers
$ f \ll 1$ such that $1 +  f \simeq 1$ and $f \simeq e^{-E/T}$.
We are here interested in the wave classical limit, that corresponds to large occupation numbers
$ f \simeq T/E \gg 1$.
The  classical wave term arises at leading order $f^3$~\cite{hep-ph/0212198,hep-ph/0412121} where
\beq F \simeq  f_1(E_1)  f_2(E_2) [ f_1(E'_1)+ f_2(E'_2)]- f_1(E'_1) f_2(E'_2)[ f_1(E_1)+ f_2(E_2)].\eeq
In this limit $\hbar$ factors cancel leaving the classical Boltzmann equation
\beq \label{eq:drho1}
\dot \rho_1 = -4\lambda^2  \int d\vec k_1d\vec k_2d\vec k'_1d\vec k'_2\,
\omega_1\, (2\pi)^{d+1}\delta(K_1+K_2-K'_1-K'_2)  \times 
\frac{\omega_1-\omega'_1}{\omega_1 \omega'_1 \omega_2 \omega'_2} \, T_1 T_2 (T_1-T_2)
\eeq
where the latter term is $\hbar^3 F$.
One can similarly compute the contribution to $\dot\rho_1$ from $11' \leftrightarrow 22'$ scatterings.
Furthermore, a $g \varphi_1^2 \varphi_2/2$ interaction among positive-energy fields $\varphi_{1,2}$
gives rise to
$2 \leftrightarrow 1 1'$ decays for $m_2 > 2 m_1$ such that
\beq \label{eq:drho1cubic}
\dot \rho_1 = - \int d\vec k_1 d \vec k'_1d\vec k'_2\,
\omega_1\, (2\pi)^{d+1}\delta(K_1+K_1'-K_2) |\mathscr{A}|^2  F
\eeq
with $\mathscr{A}= g \hbar^{1/2}$ and
\beq F =   f_1(E_1) f_2(E'_1)[1+ f_2(E_2)] 
-  f_{2}(E_2) [1+ f_1(E'_1)][1+ f_2(E'_1)] 
\simeq \frac{T_1(T_1-T_2)}{E_1E'_1}
\eeq
in the classical limit.

\bigskip

We can now repeat the computation assuming that $\varphi_2$ is a ghost.
Boltzmann equations again involve a sum over on-shell processes,
and the resonance condition among $\omega$'s now has an extra $-$ sign when a ghost is involved, see e.g.~eq.\eq{W1model}.
This is equivalent to telling that ghosts appear with negative energy in the quantum Boltzmann equations.
One can re-express the unusual (negative-energy)
kinematical integrals in terms of usual (positive-energy) ones
by rewriting each ghost wave vector  as $K_\mu = -\tilde{K}_\mu$, so that
a negative-energy particle in the initial (final) state becomes a positive-energy particle in the final (initial) state.
In the limit where each field is thermal, the Bose-Einstein distribution satisfies the identity $f(E/T)=-(1+f(- E/T))$, 
so statistical factors too match those of the positive-energy process,
up to an overall $-$ sign when an odd number of ghosts is flipped.
Let us consider some examples:
\begin{itemize}
\item
A $\varphi_1^2 \varphi_2^2$ ghost interaction allows the
kinematically open on-shell processes $12 \leftrightarrow 1'2'$ and $11'22' \leftrightarrow \emptyset $, that
become $1\tilde 2' \leftrightarrow 1 \tilde 2$ and $11' \leftrightarrow \tilde 2 \tilde 2'$.
In the classical limit one then has
$\dot \rho_1 \propto +T_1 T_2 (T_2-T_1)$  both in the ghost and the non-ghost cases.

\item 
A $\varphi_1 \varphi_2^2$ ghost interaction allows
the kinematically open on-shell process $1 2 2' \leftrightarrow  \emptyset  $, 
that becomes a $1 \leftrightarrow \tilde 2\tilde2' $ decay.
In the classical limit one then has
$\dot \rho_1\propto +T_2 (T_2-T_1)$ both in the ghost and the non-ghost cases.

\item 
A $\varphi_1^2 \varphi_2$ ghost interaction allows
the kinematically open on-shell process $ \emptyset \leftrightarrow 1 1' 2 $ that becomes a $\tilde 2 \leftrightarrow 1 1' $ decay.
In the classical limit one then has
$\dot \rho_1\propto +T_1 (T_2-T_1)$ in the non-ghost case,
that becomes $\dot \rho_1\propto -T_1 (T_2-T_1)$ in the ghost case.
\end{itemize}
The factors $F$ vanish in the thermal limit with a common temperature, $f(E) =1/(e^{E/T}-1)$.
However ghosts have $E_2<0$, so that a physical $f(E_2)\ge 0$ is obtained for
$T_2\le 0$: ghosts must have a negative temperature.\footnote{We verified that the non-equilibrium Kadanoff-Baym formalism (see e.g.~\cite{Teresi:2015bxg}) gives the same Boltzmann equations. In particular, for a ghost, the form of its two thermal Wightman propagators is exchanged with respect to positive-energy fields, so that initial-state ghosts are equivalent to final-state normal particles. 
In this formalism $f\ge 0$ because it is the expectation value of a positive number operator.

Previous literature studied possible 
thermal equilibrium thermodynamics for Lee-Wick resonances 
with negative classical energy~\cite{0902.1585,1101.5538,1306.2642}
finding contradictory results. We now see that there is no thermal equilibrium.}
We now see the key difference that arises in the presence of a ghost:
there is no thermal equilibrium at common $T$ such that the factor $F$ vanishes thanks to detailed balance,
because the two systems have opposite-sign energies and thereby temperatures.
In all cases listed above this means that the non-ghost system heats up, $\dot \rho_1 > 0$.
This sign of the heat flow agrees with our earlier considerations about increase of entropy $\dot S\ge 0$:
both $|T_1|$ and $|T_2|$ increase, as higher temperature allows for more states.
Boltzmann equations add that the energy flow rate is proportional to the coupling squared.






The purely quantum effect will be studied in section~\ref{QFT}.
We here study the classical effect, that can be isolated
as long as  the low-frequency modes excited classically $\omega \circa{<}\omega_{\rm max}$
are separated from the high-frequency modes at which the divergent
quantum effect starts giving a larger contribution to $\dot\rho_1$.
In such a case, the quantum contribution is smaller than the classical contribution
assuming a cut-off $\Lambda_{\rm UV}\circa{>} \omega_{\rm max}$.



We next compare these analytic results with numerical classical simulations in toy models,
and finally provide estimates for situations of physical interests.


%


\begin{figure}[t]
$$\includegraphics[width=0.45\textwidth]{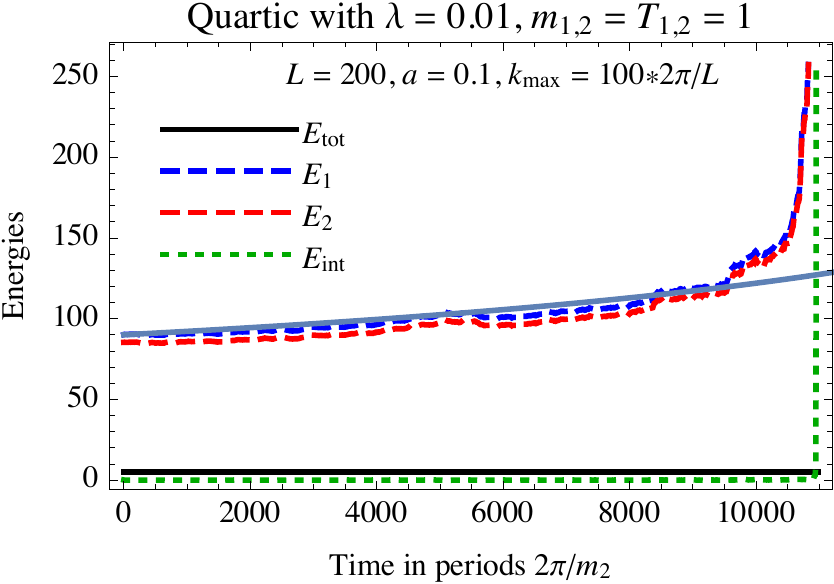}\qquad
\includegraphics[width=0.45\textwidth]{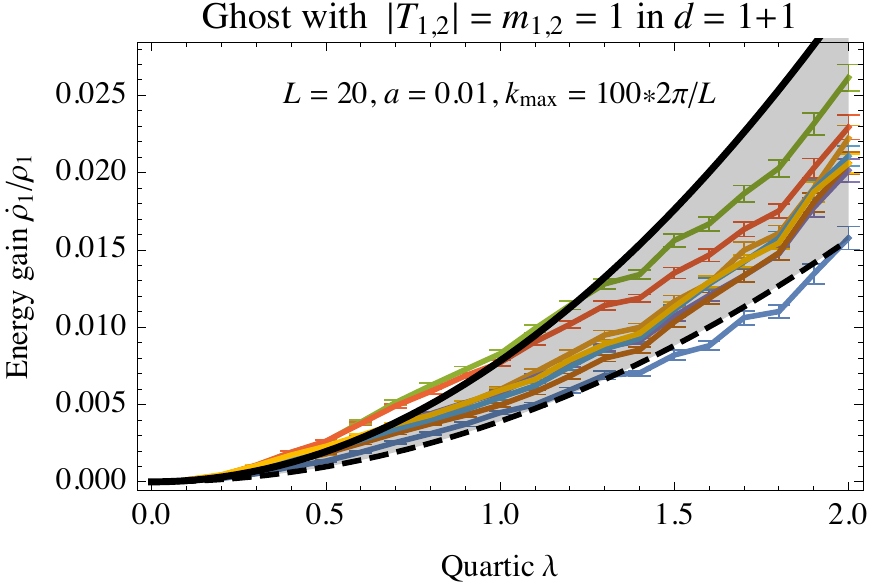}$$
\caption{\em Left: time evolution of the total energies of the normal field, of the ghost field,
of their interaction energy, of the total conserved energy. 
The continuous curve is the analytic approximation. 
Right: heat flow $d\rho_1/dt$ as function of the coupling. The data point are from lattice simulations, for
different small values of $d t$. The black curve is the analytic result; 
we also show the analytic result without the IR-divergent diagram
that might contribute in the numerics on longer time-scales (dashed curve).
\label{fig:GhostQuarticFieldEvo}}
\end{figure}

\begin{figure}
$$\includegraphics[width=0.8\textwidth]{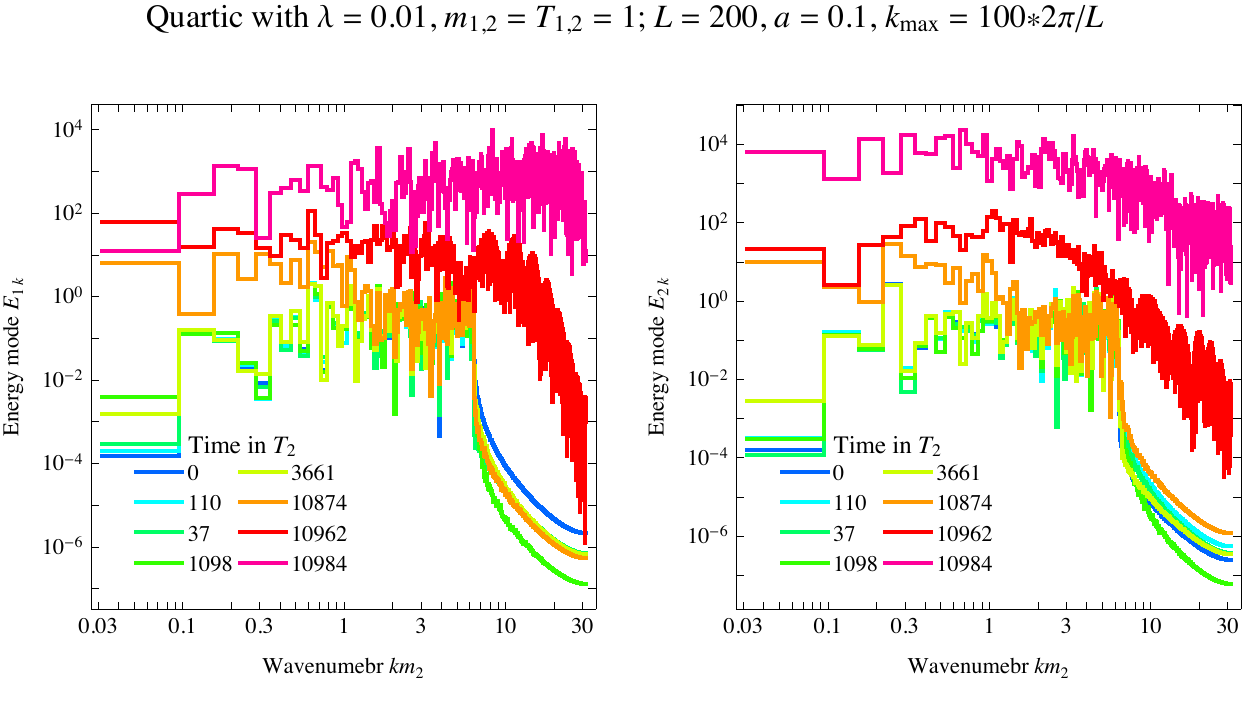}$$
\caption{\em Energy spectrum of the normal field (left) and of the ghost field (right) at some fixed times. 
\label{fig:GhostQuarticKEvo}}
\end{figure}

\subsection{Results}\label{NumericalRes}
First we simulate the classical thermalization among two positive-energy fields, finding that the
simulated rate agrees with the rate obtained from Boltzmann equations such as eq.\eq{drho1gen}.

We next consider a positive-energy field $\varphi_1$ interacting with a ghost field $\varphi_2$.
We numerically simulate their time evolution for $m_{1,2} = 1$ and $\lambda=0.01$ in $1+1$ dimensions
on a lattice with spacing $a=0.1$ and size $L=200$.
We start from a thermal-like distribution with $T_{1,2}=1$ cut at the maximal momentum $k_{\rm max} = 200\cdot 2\pi/L$.
This means that each excited mode has an initial amplitude as extracted from the thermal distribution and a random phase.
Fig.\fig{GhostQuarticFieldEvo}a shows that the system undergoes ghost run-away.
Fig.\fig{GhostQuarticKEvo} shows the energy spectra of $\varphi_1$ (left) and $\varphi_2$ (right) at some selected times.
We see that modes at higher $k$ get progressively excited: energy cascades towards the UV giving rise to the usual
black-body instability of interacting field theories (see e.g.~\cite{hep-ph/0306124,hep-ph/0410280}).  
In order to disentangle this phenomenon (that lowers $T_1$)
from ghost run-away (that increases $T_1$) we 
choose a small enough $k_{\rm max}$ such that modes around the cut-off are still negligibly excited when ghost run-away happens.

Each random initial condition with fixed temperatures produces final run-away times that differ by order one.
In order to better compare with the analytic approach, 
that predicts the average energy flow $\dot\rho_1$ between the two fields,
we run for a short time many different simulations with the same initial temperatures
and average over them.
Having assumed one spatial dimension and $m_1 = m_2$ we can analytically perform
the integrals in the Boltzmann eq.\eq{drho1},
\beq
\dot \rho_1 = \frac{ \lambda^2 T_1  T_2 (T_2-T_1)}{4\pi^2 m_{1,2}^4}
\bigg[(\ln 4-1)  + \frac12 \( 1 + \ln \frac{L m}{8 \pi}\) \bigg]
\eeq
where we added, in the second term, the contribution of  $11' \leftrightarrow 22'$ scatterings.
This process contains a logarithmic IR divergence at vanishing relative velocity between the particles,
which is typical of field theory in 1 spatial dimension.\footnote{In order to isolate the IR divergence, 
it is useful to to put the $11' \leftrightarrow 22'$
 contribution to $\dot \rho_1$ into the form
\be 
T_1 T_2 (T_2 - T_1) \, \frac{2 \lambda^2}{\pi^2} \int_{4 m^2}^\infty \!\!\! d s \int_{\sqrt{s}}^\infty \!\!\! d K_0 \frac{s^2 K_0^2}{s \sqrt{K_0^2 - s} (s - 4 m^2) (s^2 + 4 m^2 (K_0^2 -s))^2} .
\ee
In lattice simulations the IR divergence gets regulated by the finite box size $L$, so that
the lower limit of $s$ integration changes into $(2 m + 2\pi/L)^2 \simeq 4 m^2 + 8 \pi m / L$. }
Despite this aside issue, fig.\fig{GhostQuarticFieldEvo}b shows that the analytic rate agrees with the numerical rate.
We can next compute $\rho_1$ in terms of $T_1$
$$ \rho_ 1= \int E_1 \, dn_1 = T_1 \int_{2\pi/L}^{k_{\rm max}}\frac{dk_1}{2\pi}$$
and obtain a differential equation $\dot{T}_1 = \gamma T_1 T_2 (T_2-T_1) = - \dot T_2$
that can be solved
\beq 
T_1(t)  = \frac{T_{10}+T_{20}}{2}\left[1+
 \left( 1+ \frac{4T_{10} T_{20}}{(T_{10}+T_{20})^2}  e^{\gamma t (T_{10}+T_{20})^2/2} \right)^{-1/2}\right]\eeq
obtaining the average time evolution (one example is plotted in fig.\fig{GhostQuarticFieldEvo}a).


We next vary the lattice spacing, box size and number of digits used in the numerics, finding consistent results.
We also run for different values of the physical parameters; additional IR divergences arise when a field is massless.
Running for special initial conditions, such as starting from a single excited mode, $f(E) \propto \delta(E-E_0)$,
blocks or delays the ghost run-away until when enough modes can get excited, so that many resonances can happen. 




%

\medskip

Based on the above experience, we can now consider the more complicated theory of possible physical interest:
4-derivative gravity.
First, the resonances caused by cubic interactions present in 4-derivative gravity, while potentially un-safe, 
are stabilised by quartic and higher interactions, as argued at the end of section~\ref{classres}. 
Then, each resonance causes an $\mathcal{O}(1)$ energy flow
variation and the system as a whole evolves statistically, as described above.
The massive ghost present in 4-derivative gravity only has Planck-suppressed non-renormalizable interactions.
Thereby its run-away rate $\Gamma \equiv \dot \rho/\rho \sim T^3/M_{\rm Pl}^2$
is smaller than the Hubble cooling rate $H \sim T^2/M_{\rm Pl}$.
As usual, gravitational short-range interactions give negligible effects in big-bang cosmology.

Furthermore, inflation with Hubble constant $H$ roughly behaves as a thermal bath with
temperature $T \sim H$, producing a spectrum of primordial inflationary fluctuations for the graviton, its ghost, and the other fields.

In conclusion, a ghost undergoes run-away in classical field theory, 
but in 4-derivative gravity ghost run-away is negligibly slow on cosmological time-scales.




\bigskip

\section{Ghost meta-stability in quantum field theory?}\label{QFT}
We again consider a field theory with two scalars $\varphi_1$ (positive energy) and $\varphi_2$ (negative-energy ghost) in $d$ space dimensions.
We want to compute the purely quantum rate for the qualitatively new processes
where particles are emitted from the Lorentz-symmetric vacuum.
For example a  $g \varphi_1^2 \varphi_2 / 2$ interaction allows for the 3-body process
$\emptyset \leftrightarrow 1 1' 2 $.

The rates of such processes
can be obtained from the finite-temperature rates discussed in the previous section in the limit 
$T_1 \to 0^+$, $T_2\to 0^-$  and thereby $f_{1,2}\to 0$,  $1 + f_{1,2} \to 1$.
Following the discussion in section~\ref{thermalizationrate}, it is convenient to rewrite the Boltzmann equation 
in terms of positive energies $\tilde 2  \leftrightarrow 1 1'$ by defining $\tilde K_2 = - K_2$.
Since $f_2(-E_2/T_2)\to -1$ the statistical factor at zero temperature is $F \to -1$,
while it would be $F =0$ for a usual process involving only positive-energy particles.
The resulting quantum rate for the 3-body process
$\emptyset \leftrightarrow 1 1' 2$
\be 
\dot \rho_1 = \frac{\hbar^2 g^2}{2^{3 d -1} \pi^{d-1} \Gamma(d/2)^2} \frac{(m_2^2 - 4 m_1^2)^{\frac{d}{2}-1}}{m_2} \int_{m_2}^{\infty} d K_0 \,K_0 \, (K_0^2 - s)^{\frac{d}{2}-1} 
\ee 
contains a UV-divergent integral over $K_0$.


Similarly, an interaction $\lambda \varphi_1^2 \varphi_2^2/2$ allows for the 4-body process $\emptyset \leftrightarrow 1 1' 2 2' $ 
that leads to the energy flow rate
\beq
\dot \rho_1 = \int d\vec k_1d\vec k'_1 d\vec k_2  d\vec k'_2\,
E_1\, (2\pi)^{d+1}\delta(K_1+K_1'- \tilde K_2- \tilde K'_2) \frac12 |\mathscr{A}|^2.
\eeq 
By introducing $K \equiv K_1+K'_1 = \tilde K_2 + \tilde K'_2$ and $s\equiv K^2$ it becomes
\beq
\dot \rho_1 = \frac{\hbar^3 \lambda^2}{2^{5d-3} \pi^{\frac{3 d}{2} -1} \Gamma(d/2)^3} \int_{4 m^2}^{\infty} d s \, \frac{(s-4 m^2)^{d-2}}{s} \int_{\sqrt{s}}^{\infty} d K_0 \,K_0 \, (K_0^2 - s)^{\frac{d}{2}-1}.
\eeq 
Again, the integral over $K_0$ is UV-divergent.

This new divergence arises because, unlike in the thermal case, 
the vacuum initial state $\emptyset$ is now Lorentz-invariant
so that the final state too must be the same in all frames.
This is why the rate contains a $dK_0$  integral over the non-compact Lorentz group.

%
%


\smallskip

This is the same divergent  `boost'  integral discussed by~\cite{Zeldovich:1974py,Kobzarev:1974cp} (and more recently by~\cite{1107.0956}).
These early studies of vacuum decay considered a theory containing a scalar with positive kinetic energy 
(no ghost) and assumed that its potential $V$
contains a local minimum e.g.\ with $V=0$ and a deeper minimum with $V<0$.
The vacuum decay bubble with mass $m=0$ can appear with any initial velocity, giving rise to the
divergent Lorentz integral~\cite{Zeldovich:1974py,Kobzarev:1974cp}.
Furthermore, by e.g.\ increasing its radius one obtains field configurations with generic $m_2<0$,
that thereby have negative energy  with $K_2 = (m_2,\vec 0)$.
Such ghost configurations can be emitted from the vacuum together with one particle with positive energy $K_1 = (m_1,\vec 0)$,
for $m_1+m_2=0$.
Due to relativistic invariance, this process happens with the same amplitude for
arbitrarily boosted $K_2$ and $K_1$, giving rise to a divergent $dK_0$ integral over boosts~\cite{1107.0956}.

\smallskip

One then wonders if  both $K$-instability (ghosts) and $V$-instability (vacuum tunnelling) proceed
with infinite rate, in contradiction with our usual understanding of vacuum tunnelling
as exponentially slow~\cite{1107.0956}.


\smallskip

In the case of $V$-instability, Coleman~\cite{Coleman:1977py} and more recently~\cite{1109.3422}
interpreted the Lorentz boost divergence as emission of lots of extra quanta i.e.\
that the naive perturbative computation is not
expanding the path integral around the right saddle point.\footnote{Other authors regulate the boost divergence through cosmology adding a Lorentz-breaking or non-local cut-off~\cite{Cline:2003gs,hep-th/0505265,1202.1239}.}
These authors argue that vacuum tunnelling must instead be
computed expanding around a Lorentz-invariant `bounce' configuration, such that
an integral over the Lorentz group is not needed because it
would be an over-counting of the same configuration.
Accepting this argument, the WKB approximation allows to find the desired configuration
as the `bounce' instanton that minimises an effective Euclidean action.
The `bounce' is the solution to the
scalar field equations that only depends on the Euclidean $r^2_{\rm E} = x^2+y^2 + z^2 + (it)^2$
(the Euclidean Lorentz group is compact) and has the desired boundary conditions: false vacuum at $r\to \infty$ and
over the barrier at $r\to 0$:
\beq \label{eq:Colemanbou}
\left\{\begin{array}{ll}\varphi_i (r) = 0 & \qquad\hbox{as $r\to \infty$, false vacuum} \\
\dot\varphi_i(r) =0 & \qquad\hbox{as $r\to 0$, true vacuum}
\end{array}\right.  .
\eeq
The resulting vacuum decay rate is exponentially suppressed by the coupling, $ e^{-{\cal O}(1)/\lambda}$.

\smallskip

In the ghost case, we do not have a similarly simple formulation
nor a positive Euclidean action.
Unless a suitable continuation is found, a brute-force computation is needed
to establish if the ghost decay rate is exponentially suppressed
(restricting the action to Lorentz-invariant field configurations removes
field-theory resonances but leads to $r$-depended frequencies).
We speculate that, if the vacuum decay rate will turn out to be exponentially suppressed,
the difficulties that seem to hinder unitarity and/or renormalizability of Minkowskian theories with ghosts~(see e.g.~\cite{Cutkosky:1969fq,Nakanishi:1971jj,Lee:1971ix,Boulware:1983vw})
will turn out to be similarly suppressed by similar factors.

\section{Conclusions}\label{concl}
Systems containing positive kinetic energy $K_1$ interacting with negative kinetic energy $K_2$ can 
undergo a run-away where the total energy $E = K_1 + K_2 + V$ is constant while $|K_i|\to \infty$.
Thereby negative kinetic energy is considered as unphysical and dubbed `ghost'.
We explored the possibility that negative kinetic energy can be physically acceptable
because meta-stable up to cosmological times, similarly to
negative potential energy.
In order to exclude this possibility we started from the simplest limit (classical mechanics),
but we found that a weakly-interacting ghost behaves almost as well as a free ghost:
\begin{itemize}
\item In section~\ref{class} we found that {\bf ghosts are meta-stable in classical mechanics}.
Recent numerical studies rediscovered that, in some cases, energies of individual degrees of freedom surprisingly 
remain confined to a finite region despite that no constant of motion imposes such lock-down.
Ghost meta-stability is understood  using the same mathematical techniques developed in the past centuries to study if 
multi-body systems like the solar system 
are stable up to cosmological times despite that individual planets can acquire enough energy to escape.
One  `diagonalises' the classical Hamiltonian by performing a perturbative expansion
around the limit where each degree of freedom undergoes periodic motion with pulsation $\omega_i$.
Technically, this means finding a canonical transformation to action-angle variables such that the Hamiltonian
does not depend on angle variables.
If interactions are strong, 
outside the convergence radius of the perturbative series, motion is chaotic, planets escape and ghosts run-away.
If interactions are weak the perturbative
series is convergent: planets undergo quasi-periodic motion with epicycles, and ghosts are stable.
The dimension-less expansion parameter is the energy in the interaction term divided by the energy in the
free quadratic part of the Hamiltonian.
Ghost lock-down within finite regions of phase space is understood as due to hidden quasi-constants of motion
present in almost generic theories at weak coupling.
Extending towards infinite time reveals an exponentially suppressed run-away rate, that we controlled in some model.

Actually  some physical systems are meta-stable ghosts, such as 
asteroids around the Lagrangian point 4 (appendix~\ref{L4}) or electrons
in magnetic fields plus a destabilising radial force (appendix~\ref{eB}).

\item However the perturbative series contains terms proportional to $1/(N_1 \omega_1 - N_2 \omega_2)$
where $N_i$ are integers that grow at higher orders.  
One can thereby encounter resonances where such terms are large or divergent.
The most dangerous case arises at leading order $N_{1,2}=1$ when $\omega_1=\omega_2$.
We studied what happens using {\em resonant normal forms}: some interactions lead to ghost run-away,
others only to order-one violations of hidden quasi-constant of motion.
We argued that the latter situation seems quite generic in the presence of multiple interactions.

\end{itemize}
In order to exclude a ghost, we then moved to less simple limits:
\begin{itemize}

\item In section~\ref{QM} we argued that {\bf ghosts are meta-stable in quantum mechanics}.
We first performed a brute-force computation in our toy model.
Wave-functions with no nodes (the ground-like state with lowest positive energy and highest negative energy)
get exponentially-suppressed away from the origin even into the dangerous new region that leads to ghost run-away
(large $|K_i|$ and small $K_1+K_2$). 
The ghost run-away time is thereby exponentially suppressed at small coupling 
analogously to usual tunnelling.
In general, tunnelling can be approximated in the semi-classical limit,
that inherits the good properties of ghosts in classical mechanics.
We could however not generalise the WKB simple formula to the ghost case.


\item In section~\ref{CFT} we studied {\bf classical field theory}.
The infinite number of degrees of freedom give rise to new phenomena.
One is the black-body problem of interacting classical field theories, that complicates our study.
More relevant for us is the presence of an infinite number of modes with different frequencies 
and thereby an infinite number of resonances,
that correspond to the usual on-shell decays and scatterings.
Each resonance is potentially deadly in the presence of ghosts.
By expanding examples of local interactions in terms of momentum modes
we found specific resonances that do not immediately lead to run-aways,
but only to partial loss of hidden constants of motion.
Nevertheless we argued that the infinite number of resonances makes ghosts
unprotected in the continuum limit.
Based on general entropy arguments we found that there is no thermal state when
a system with positive temperature $T_1>0$ interacts with a ghost system with negative $T_2<0$:
heat keeps flowing such that both $|T_{1,2}|$ increase up to infinity.
By writing Boltzmann equations in specific models we computed the rate of such process,
finding that it is quadratic in the couplings, rather than non-perturbatively suppressed.
We validated this finding by evolving classical field theories on appropriate lattice discretizations.
In principle both our analytic understanding and the numerics might have missed hidden
properties that keep ghosts stable, but various checks do not find evidence in this sense.

We next considered the case of 4-derivative gravity ---
a renormalizable theory of gravity containing a spin-2 field with negative kinetic energy
and gravitational interactions only --- finding that the ghost
run-away time is negligible on cosmological time-scales.

\end{itemize}
In order to exclude such ghost, we finally considered the theory currently considered as fundamental.
\begin{itemize}
\item In section~\ref{QFT} we considered Relativistic Quantum Field Theory in the presence of a ghost.
Since the initial vacuum state is Lorentz invariant (unlike a thermal state),
the naive tree-level vacuum decay rate contains a divergent integral over the non-compact
Lorentz group, that describes an arbitrary boost of the same final state.
We recalled that this same problem was encountered in early computations of vacuum decay
due to potential instability: even in the absence of a ghost, 
negative potential energy gives rise to field configurations that behave as a ghost.
Using WKB Euclidean techniques Coleman argued that the vacuum decay rate is finite and exponentially suppressed.
We could not extend such tecniques to the case of ghost instability,
so we do not know if it is fast (thereby ruling out theories containing ghosts) or exponentially suppressed
at small couplings.
\end{itemize}
It will be important to fully clarify if negative kinetic energy can be meta-stable up to cosmological time-scales,
as the negative-energy quantization of 4-derivative gravity would provide a renormalizable theory of quantum gravity.

\small

\subsubsection*{Acknowledgments}
This work was supported by the ERC grant 669668 NEO-NAT.
We thank Gia Dvali, Enore Guadagnini, Riccardo Rattazzi, Michele Redi for discussions.
 

\appendix

\section{Physical systems described by ghosts}\label{mainstream}

\subsection{Asteroids around the Lagrangian point $\Lag_4$}\label{L4}
Let us consider an asteroid with negligible mass around Lagrangian point $\Lag_4$ of the
Sun/Jupiter system.
The quadratic part of the asteroid Hamiltonian contains a negative frequency (see e.g.~\cite{Giorgilli});
we next show that it is a ghost degree of freedom (negative kinetic energy).

The Hamiltonian of a free particle with mass $m$ in a reference frame rotating with
angular velocity $\omega$ around the $z$ axis is $H_{\rm free} = \vec p\,{}^2/2m +\omega (y p_x-x p_y)$.
We compute the  Hamiltonian of an asteroid in the center-of-mass frame of the Sun/Jupiter system,
where the Sun is fixed at $\vec x_S = (-\mu,0,0)$ and Jupiter at $\vec x_J=(1-\mu,0,0)$. 
In suitable units their masses are $M_J = \mu$ and
$M_S = 1-\mu$. 
The asteroid Hamiltonian in the $x,y$ plane is
\beq H = \frac{\vec p\,^2}{2} + y p_x-x p_y - \frac{M_S}{|\vec x - \vec x_S|} - \frac{M_J }{|\vec x - \vec x_J|}.\eeq
The momentum $p$ has a possible stationary point at $z=0$, $p_x = -y$ and $p_y=x$.
Inserting this in $H$ gives an effective potential with stationary points along the $x$ axis,
as well as at the $\Lag_4$ points $x = \frac{1}{2}-\mu$ and $y=\pm\sqrt{3}/2$.
Interesting motion happens along the $xy$ plane and we can ignore motion along the $z$ axis.

Expanding $H$ around $\Lag_4$ gives, at quadratic order,
\beq H_2 = \frac{p_x^2 + p_y^2}{2} + y p_x - x p_y + \frac{x^2}{8} - \frac{5 y^2}{8} + \frac{\sqrt{27}}{4} (2\mu-1) xy.\eeq
Writing such quadratic part of the Hamiltonian as $H_2=\frac12 v_i  \hat{H}_{ij} v_j $ where $v \equiv(x,y,p_x,p_y)$, the Hamilton equations are $\dot v=\hat{J}  \hat{H} v$, where 
\bal
\hat{J}&= 
\left( 
\begin{array}{cc}
 {0}_{2\times 2} & \One_{2\times 2}\\
 - \One_{2\times 2} &  {0}_{2\times 2} \\
\end{array} 
\right) 
\eal
is the symplectic invariant tensor.
The eigenvalues of $\hat{J} \hat{H}$ give the frequencies of the normal modes.
Since $\hat{H}$ is real and symmetric,
if $\lambda$ is an eigenvalue, then  $-\lambda, \lambda^*,-\lambda^*$ too are eigenvalues.
Thus we can write the four eigenvalues as $(i \omega_1,-i \omega_1,i \omega_2,-i \omega_2)$.
We are interested in the case where $\omega_{1,2}$ are real so that the solutions to the equations of motions for the linearized Hamiltonian $H_2$ are stable oscillations rather than exponential tachyonic solutions (a free $2\times 2$ Hamiltonian has eigenvalues $\pm i \omega$, such that $e^{\pm i \omega t}$ solutions
give sine and cosine).
Restricting without loss of generality to the interval $0<\mu<1/2$, the eigenvalues are imaginary for $0<\mu<\mu_{\rm Routh}$ where $\mu_{\rm Routh}=\frac12 (1-\sqrt{23/27})\approx 3.9 \times 10^{-2}$ (the Jupiter-Sun system corresponds to  $\mu \approx 0.95  \times 10^{-3}$).
One finds the frequencies
\beq
\omega_{1,2} = \sqrt{\frac{1\pm r}2}  \quad \textrm{where} \quad r=\sqrt{1- 27 \mu (1-\mu)} \,.
\eeq
$H_2$ is not positive definite, signalling the presence of a tachyon (negative potential energy)
and/or of a ghost (negative kinetic energy).
To clarify, we identify the normal modes by bringing $H_2$ to normal form through a linear change of variables $v= \hat{N} v'$, 
where $\hat{N}$ must be real and symplectic (i.e. $\hat{N}^T \hat{J} \hat{N}=\hat{J}$) in order to preserve the Hamiltonian structure of the equations of motion.
The needed Sp(4) rotation is~\cite{russians}
\beq
\hat{N}=\left( \frac{\mathrm{Re}(z_1)}{\sqrt{|c_1 |}}, \frac{\mathrm{Re}(z_2)}{\sqrt{|c_2 |}}, \mathrm{sign}(c_1) \frac{\mathrm{Im}(z_1)}{\sqrt{|c_1 |}}, \mathrm{sign}(c_2)  \frac{\mathrm{Im}(z_2)}{\sqrt{|c_2 |}}\right)
\eeq
where $z_{j}$ are the complex eigenvectors of $\hat{J} \hat{H}$ corresponding to the eigenvalues $+i \omega_{j}$
(the opposite convention is also applicable)
and $c_j=\mathrm{Re}(z_j)^T \hat{J} \, \mathrm{Im}(z_j)$.
Writing $v'=(q_1,q_2,p_1,p_2)$, the diagonalised Hamiltonian is
\beq 
H_2 = \omega_1 \frac{p_1^2 + q_1^2}{2} - \omega_2 \frac{p_2^2 + q_2^2}{2} .
\eeq
As expected $H_2$ is not positive definite, and the ghost is $q_2,p_2$.
At linear order the system is stable, because the two oscillators do not interact.
At higher order the ghost couples to the normal oscillator and one might expect quick run-away.
Still, asteroids remain close to $\Lag_4$ for exponentially long time~\cite{Giorgilli}.

One can maybe more intuitively see how a positive-energy particle written in a rotating frame
becomes a ghost in the Lagrangian formalism.
A free particle is described by
$L = (\dot x^2+\dot y^2)/2 +\omega (x\dot y - y \dot x) + \omega^2 (x^2+y^2)/2$.
The second term is the Coriolis force.
The third term is the centrifugal force: kinetic energy become a potential term. 
Thereby  extra potential terms (such as gravity) can modify the kinetic term, giving rise to a ghost.



\subsection{Charged particle in a magnetic field}\label{eB}
The Hamiltonian of a non-relativistic particle with mass $m$ and
electric charge $e$ in a constant magnetic field $\vec B =(0,0,B_z)$ described by the vector potential 
$\vec{A} =\vec{B}\times\vec{r}/2$ is
\beq H_0 = \frac{(\vec p - e \vec A)^2}{2m}+e\varphi = 
\frac{\vec p\,^2}{2m}+\omega_B (y p_x -x p_y) + \frac{m}{2} \omega_B^2 (x^2+y^2).
\eeq
The first two terms are equal to the Hamiltonian of a free particle written in a frame rotating with cyclotron frequency
$\omega_B = \sfrac{e B_z}{2m}$.
The equations of motion give $m\dot{\vec{x}} =\vec p-e\vec A$, showing that the magnetic force does
not affect energy.
We add to $H_0$ a {\em de}stabilising potential $\delta H= -m{\omega^2_0}(x^2+y^2)/2$, $H=H_0+\delta H$.
The eigenvalues of $\hat{J}\hat{H} $ are $\pm i \omega_\pm$ with
\beq \omega_\pm=\omega_B \pm \delta \omega  \quad \textrm{where} \quad \delta \omega=\sqrt{\omega_B^2-\omega_0^2} \,. \eeq
For $0<\omega_0^2 <\omega_B^2$ one has $\omega_+ > \omega_->0$ and, diagonalising $H$ via a canonical transformation
\beq 
H= \omega_+ \frac{p_+^2 + q_+^2}{2} - \omega_- \frac{p_-^2 + q_-^2}{2},
\eeq
shows that the $-$ mode is a ghost.
The two pulsations $\omega_\pm$ become degenerate for $\omega_0^2 = \omega_B^2$
(in this limit one has the same $H$ as a free particle seen from a rotating frame),
and tachyons appear for $\omega_0^2 > \omega_B^2$.


\section{Resonant form for overlapping resonances} \label{sec:overlapping}
In this appendix, we repeat the argument of section~\ref{fieldKAM} for the case of multiple resonances.
For the system considered in section~\ref{fieldKAM}, this can happen if and only if all frequencies are approximately equal, $\omega$. Therefore, three resonant combinations are now present:
\begin{align} 
4 \Q_s \equiv \Q_1' + \Q_2' + \Q_3' + \Q_4',\qquad
4 \Q_t \equiv \Q_1' - \Q_2' - \Q_3' + \Q_4',\qquad
4 \Q_u \equiv \Q_1' - \Q_2' + \Q_3' - \Q_4' .
\end{align}
The corresponding resonant form is
\begin{align}\label{eq:resonant_allequal}
H &\simeq \omega (\P_1' + \P_2 ' - \P_3' - \P_4') + \frac{\epsilon}{4} \bigg[  \P_1' \P_3' + \P_1' \P_4' + \P_2' \P_3' + \P_2' \P_4' + \sqrt{\P_1' \P_2' \P_3' \P_4'} \( \cos 4 \Q_s + \cos 4 \Q_t +\cos 4 \Q_u \) \bigg]
\end{align}
The only quasi-integral of motion (in addition to $H$) is $\mathcal{E} \equiv \P_1' + \P_2' - \P_3' - \P_4'$. The Hamiltonian of the extra-system can be easily obtained from eq.~\eqref{eq:resonant_allequal} and, recalling that the combination $\mathcal{E}$ is approximately constant, is found to be bounded
(this can be seen by  noticing that the absolute value of the oscillatory term in the square brackets is smaller than $4 \sqrt{\P_1'\P_2'\P_3'\P_4'}$ and using twice the inequality $2 \sqrt{xy} < x+y$ between arithmetic and geometric means).

\section{Classical lattice simulations}\label{lattice}
We consider the Lagrangian of eq.\eq{Lagphi12} with a $\lambda \varphi_1^2 \varphi_2^2/2$ interaction
in 1+1 dimensions with coordinates $(x_0,x_1)$. 
We express all dimensionful quantities in units of the ghost mass $m_2$ by introducing the dimensionless coordinates $t\equiv m_2  x_0$ and $x\equiv m_2  x_1$, as well as the dimensionless parameters $\kappa \equiv m_1^2/m_2^2$ and $\bar \lambda \equiv \lambda/m_2^2$.
Then we obtain the dimensionless Lagrangian
\be
\frac{{\Lag}}{m_2^2} \equiv \bar {\Lag} =\frac12 \left[ ( \dot{\varphi_1}^2-\varphi_1^{\prime 2}-  \kappa \,  \varphi_1^2) - 
(\dot{\varphi_2}^2-\varphi_2^{\prime 2}- \varphi_2^2)- \bar \lambda \varphi_1^2 \varphi_2^2 \right] \,.
\ee
The equations of motion are
\beq
 \label{eoms}
\left\{\begin{array}{l}
\ddot \varphi_1 - \varphi_1''+\varphi_1(\kappa+ \bar \lambda \varphi_2^2)=0 \nn\\
\ddot \varphi_2 - \varphi_2''+\varphi_2(1 - \bar \lambda \varphi_1^2)=0 \,.
\end{array}\right.
\eeq
These nonlinear 2nd-order hyperbolic partial differential equations can be solved with finite-difference lattice methods.
For a $\varphi^4$ theory this has been done in 1+1 \cite{hep-ph/0306124} and 3+1 \cite{hep-ph/0410280} dimensions using a light cone lattice (namely, a square lattice in $x\pm t$ coordinates)
and an exactly conserved energy on the lattice.
We generalise this procedure to two fields.
This is non-trivial, as one needs to achieve energy conservation around cut-off scales while avoiding
choices that lead to impractically complicated discretised field equations.

The continuum Hamilton density is $\bar {\cal H} =\frac12 \left[ ( \pi_1^2+\varphi_1^{\prime 2}+  \kappa \,  \varphi_1^2) - 
(\pi_2^2+\varphi_2^{\prime 2}+ \varphi_2^2)+ \bar \lambda \varphi_1^2 \varphi_2^2 \right]$ where $\pi_i = \dot \varphi_i$.
We introduce two lattice Hamilton densities
\be
{\cal H}_\pm= 
\frac12  \left[ (\pi_{1\pm}^2 + \varphi^{\prime \, 2}_{1\pm} + \kappa \, \varphi^{2}_{1\pm} )
-(\pi_{2\pm}^2 + \varphi^{\prime \, 2}_{2\pm} +  \varphi^{2}_{2\pm} )
+\bar \lambda [\varphi_1^{2}  \varphi_2^{2}]_\pm \right]
\ee
where we defined
\bal
\pi_{i\pm}^2 &= \left( \frac{ 2 \varphi_i(x,t_\pm) - \varphi_i(x_-,t) - \varphi_i(x_+,t) }{2 a} \right)^2
\nn \\
 \varphi^{\prime \, 2}_{i\pm} &= \left( \frac{ \varphi_i(x_-,t) - \varphi_i(x_+,t) }{2 a} \right)^2
\nn \\
  \varphi^{2}_{i\pm} &= \frac{ 2 \varphi_i(x,t_\pm)^2 + \varphi_i(x_-,t)^2 + \varphi_i(x_+,t)^2 }{4} 
  \nn \\
 [\varphi_1^{2}  \varphi_2^{2} ]_\pm&= \frac{ \varphi_1 (x_-,t) \varphi_2 (x_-,t) + \varphi_1 (x_+,t) \varphi_2 (x_+,t) }{2} \varphi_1 (x,t_\pm)  \varphi_2(x,t_\pm).
\eal
Here, $a$ is the dimensionless lattice distance and we abbreviated $x_\pm = x \pm a$ and $t_\pm = t \pm a$.
In the continuum limit $\pi_{i\pm } \to \dot \varphi_i$.
The definition $[\varphi_1^{2}  \varphi_2^{2} ]_\pm$ of the lattice interaction term 
significantly simplifies equations compared to the naive interaction term $\varphi_{1\pm}^{2} \varphi_{2\pm}^{2}$.
In the continuum limit, ${\cal H}_+$ and ${\cal H}_-$ both approach the continuum Hamilton density:
$
\lim_{a \to 0} {\cal H}_\pm = \bar {\cal H} 
$.
Their difference can be expressed as
\be
{\cal H}_+ - {\cal H}_- = \frac{\varphi_1(x,t_+)-\varphi_1(x,t_-)}{2 a^2} \, Q_{1} - \frac{{\varphi_2}(x,t_+)-{\varphi_2}(x,t_-)}{2 a^2} \, Q_{2} 
\ee
where
\bal
Q_{1} &=
\[ \varphi_1(x,t_+)+ \varphi_1(x,t_-)\] (1+ \kappa {a^2}/{2})  -[\varphi_{1}]_a(x,t)  
 +\frac{\bar \lambda a^2}{4}\[ {\varphi_2}(x,t_+)+ {\varphi_2}(x,t_-)\] [\varphi_1 {\varphi_2}]_a(x,t) 
\nn\\
Q_{2} &=
\[ {\varphi_2}(x,t_+)+ {\varphi_2}(x,t_-)\] (1+ {a^2}/{2}) - [{\varphi_2}]_a(x,t)
  -\frac{\bar \lambda a^2}{4}\[ \varphi_1(x,t_+)+ \varphi_1(x,t_-)\] [\varphi_1 {\varphi_2}]_a(x,t)  
\eal
and we defined
\bal
[\varphi_{i}]_a(x,t)&=\varphi_i(x_-,t)+\varphi_i(x_+,t) 
\nn \\
[\varphi_1 {\varphi_2}]_a(x,t)&=\varphi_1 (x_-,t) {\varphi_2} (x_-,t) + \varphi_1 (x_+,t) {\varphi_2} (x_+,t) .
\eal
Energy is exactly conserved on the lattice if $Q_{1}=Q_{2}=0$.
In the continuum limit, this condition becomes the equations of motion in eq.~(\ref{eoms}):
\bal 
\ddot \varphi_1 - \varphi_1''+\varphi_1(\kappa+ \bar \lambda \varphi_2^2)&=
-a^2\left[ \frac\kappa2 \ddot \varphi_1 +\frac{\ddddot \varphi_1 - \varphi_1''''}{12} + \frac{\bar \lambda}{2} \varphi_2 \left( (\varphi_1 \varphi_2)''+ \varphi_1 \ddot \varphi_2 \right)  \right] 
+ {\cal O} (a^4)
\nn\\
\ddot \varphi_2 - \varphi_2''+\varphi_2(1 - \bar \lambda \varphi_1^2)&=
-a^2\left[ \frac12 \ddot \varphi_2  +\frac{\ddddot \varphi_2 - \varphi_2''''}{12} - \frac{\bar \lambda}{2} \varphi_1 \left( (\varphi_1 \varphi_2)''+ \ddot \varphi_1  \varphi_2 \right)  \right] 
+ {\cal O} (a^4) \,.
\eal
So, by imposing $Q_{1}=Q_{2}=0$ and
solving for $\varphi_1(x,t_+)$ and ${\varphi_2}(x,t_+)$
we get discretised equations of motion that exactly conserve energy:
\bal
\varphi_1(x,t_+)&=-\varphi_1(x,t_-)
+\frac{\left(1+ {a^2}/{2}\right) [{\varphi_1}]_a(x,t)- \left({\bar \lambda a^2}/{4}\right)\ [{\varphi_2}]_a(x,t) \ [\varphi_1 {\varphi_2}]_a(x,t) }{\left(1+{a^2}/{2}\right)\left(1+\kappa {a^2}/{2}\right)+ \left({\bar \lambda a^2}/{4}\right)^2 [\varphi_1 {\varphi_2}]^2_a(x,t)}
\nn \\
{\varphi_2}(x,t_+)&=-{\varphi_2}(x,t_-)
+\frac{\left(1+ \kappa {a^2}/{2}\right) [{\varphi_2}]_a(x,t)+ \left({\bar \lambda a^2}/{4}\right)\ [{\varphi_1}]_a(x,t) \ [\varphi_1 {\varphi_2}]_a(x,t) }{\left(1+{a^2}/{2}\right)\left(1+\kappa {a^2}/{2}\right)+ \left({\bar \lambda a^2}/{4}\right)^2 [\varphi_1 {\varphi_2}]^2_a(x,t)} \,.
\eal
For zero interaction $\lambda=0$ the energies of $\varphi_1$ and $\varphi_2$ are separately exactly conserved.
The method can be extended to cubic interactions.

\end{document}